\documentclass[a4paper,11pt]{article}
\usepackage{jcappub} 
\usepackage{caption}
\usepackage{subcaption}

\usepackage{lineno}

\title{\boldmath  Constraints on Axion-Like Particles from Ultra-High-Energy Observations of 3HWC J1908+063 with HAWC }   
\usepackage{xcolor} 
\usepackage{seqsplit}
\usepackage{listings}
\usepackage{ulem}

\author[a,*]{R.~Alfaro}
\author[b]{C.~Alvarez}
\author[c]{A.~Andrés}
\author[c]{E.~Anita-Rangel}
\author[d]{M.~Araya}
\author[e]{J.C.~Arteaga-Velázquez}
\author[c]{D.~Avila Rojas}
\author[f]{H.A.~Ayala Solares}
\author[g]{R.~Babu}
\author[f]{P.~Bangale}
\author[a]{E.~Belmont-Moreno}
\author[c]{A.~Bernal}
\author[b]{K.S.~Caballero-Mora}
\author[c]{T.~Capistrán}
\author[h]{A.~Carramiñana}
\author[c]{F.~Carreón}
\author[i]{S.~Casanova}
\author[e]{U.~Cotti}
\author[j]{J.~Cotzomi}
\author[k]{E.~De la Fuente}
\author[l]{P.~Desiati}
\author[m]{N.~Di Lalla}
\author[h]{R.~Diaz Hernandez}
\author[l]{M.A.~DuVernois}
\author[l]{J.C.~Díaz-Vélez}
\author[g]{T.~Ergin}
\author[a]{C.~Espinoza}
\author[c]{N.~Fraija}
\author[c]{S.~Fraija}
\author[n]{J.A.~García-González}
\author[c]{F.~Garfias}
\author[o]{N.~Ghosh}
\author[a]{A.~Gonzalez Muñoz}
\author[c]{M.M.~González}
\author[e]{J.A.~González}
\author[p]{J.A.~Goodman}
\author[q]{J.~Gyeong}
\author[r]{J.P.~Harding}
\author[ab,*]{S.~Hernández-Cadena}
\author[g]{I.~Herzog}
\author[p]{D.~Huang}
\author[b]{F.~Hueyotl-Zahuantitla}
\author[c]{A.~Iriarte}
\author[s]{S.~Kaufmann}
\author[t]{D.~Kieda}
\author[u]{A.~Lara}
\author[c]{W.H.~Lee}
\author[v]{J.~Lee}
\author[a]{H.~León Vargas}
\author[c]{A.L.~Longinotti}
\author[s]{G.~Luis-Raya}
\author[r]{K.~Malone}
\author[j]{O.~Martinez}
\author[w]{J.~Martínez-Castro}
\author[x]{H.~Martínez-Huerta}
\author[y]{J.A.~Matthews}
\author[z]{P.~Miranda-Romagnoli}
\author[c]{P.E.~Mirón-Enriquez}
\author[e]{J.A.~Morales-Soto}
\author[j]{E.~Moreno}
\author[f]{M.~Mostafá}
\author[o]{M.~Najafi}
\author[i]{A.~Nayerhoda}
\author[aa]{L.~Nellen}
\author[g]{M.U.~Nisa}
\author[z]{R.~Noriega-Papaqui}
\author[m]{N.~Omodei}
\author[j]{E.~Ponce}
\author[a]{Y.~Pérez Araujo}
\author[s]{E.G.~Pérez-Pérez}
\author[a,*]{A. Pratts}
\author[q]{C.D.~Rho}
\author[e]{A.~Rodriguez Parra}
\author[h]{D.~Rosa-González}
\author[r]{M.~Roth}
\author[a]{A.~Sandoval}
\author[p]{M.~Schneider}
\author[a]{J.~Serna-Franco}
\author[p]{A.J.~Smith}
\author[v]{Y.~Son}
\author[t]{R.W.~Springer}
\author[s]{O.~Tibolla}
\author[g]{K.~Tollefson}
\author[h]{I.~Torres}
\author[ab]{R.~Torres-Escobedo}
\author[j]{E.~Varela}
\author[j]{L.~Villaseñor}
\author[ac]{X.~Wang}
\author[ac]{Z.~Wang}
\author[v]{I.J.~Watson}
\author[l]{H.~Wu}
\author[ad]{S.~Yu}
\author[i]{X.~Zhang}
\author[ab]{H.~Zhou}
\author[e]{C.~de León}

\affiliation[a]{Instituto de Física, Universidad Nacional Autónoma de México, Ciudad de México, México}
\affiliation[b]{Universidad Autónoma de Chiapas, Tuxtla Gutiérrez, Chiapas, México}
\affiliation[c]{Instituto de Astronomía, Universidad Nacional Autónoma de México, Ciudad de México, México}
\affiliation[d]{Universidad de Costa Rica, San José 2060, Costa Rica}
\affiliation[e]{Universidad Michoacana de San Nicolás de Hidalgo, Morelia, México}
\affiliation[f]{Temple University, Department of Physics, 1925 N. 12th Street, Philadelphia, PA 19122, USA}
\affiliation[g]{Department of Physics and Astronomy, Michigan State University, East Lansing, MI, USA}
\affiliation[h]{Instituto Nacional de Astrofísica, Óptica y Electrónica, 72000 Puebla, México}
\affiliation[i]{Institute of Nuclear Physics Polish Academy of Sciences, PL-31342 Krakow, Poland}
\affiliation[j]{Facultad de Ciencias Físico Matemáticas, Benemérita Universidad Autónoma de Puebla, Puebla, México}
\affiliation[k]{Departamento de Física, Centro Universitario de Ciencias Exactas e Ingenierías, Universidad de Guadalajara, Guadalajara, México}
\affiliation[l]{Dept. of Physics and Wisconsin IceCube Particle Astrophysics Center, University of Wisconsin--Madison, Madison, WI, USA}
\affiliation[m]{Department of Physics, Stanford University, Stanford, CA 94305--4060, USA}
\affiliation[n]{Tecnologico de Monterrey, Escuela de Ingeniería y Ciencias, Ave. Eugenio Garza Sada 2501, Monterrey, N.L., México, 64849}
\affiliation[o]{Department of Physics, Michigan Technological University, Houghton, MI, USA}
\affiliation[p]{Department of Physics, University of Maryland, College Park, MD, USA}
\affiliation[q]{Department of Physics, Sungkyunkwan University, Suwon 16419, South Korea}
\affiliation[r]{Los Alamos National Laboratory, Los Alamos, NM, USA}
\affiliation[s]{Universidad Politécnica de Pachuca, Pachuca, Hgo, México}
\affiliation[t]{Department of Physics and Astronomy, University of Utah, Salt Lake City, UT, USA}
\affiliation[u]{Instituto de Geofísica, Universidad Nacional Autónoma de México, Ciudad de México, México}
\affiliation[v]{University of Seoul, Seoul, Republic of Korea}
\affiliation[w]{Centro de Investigación en Computación, Instituto Politécnico Nacional, México City, México}
\affiliation[x]{Universidad de Monterrey, San Pedro Garza García, Nuevo León, México}
\affiliation[y]{Dept of Physics and Astronomy, University of New Mexico, Albuquerque, NM, USA}
\affiliation[z]{Universidad Autónoma del Estado de Hidalgo, Pachuca, México}
\affiliation[aa]{Instituto de Ciencias Nucleares, Universidad Nacional Autónoma de México, Ciudad de México, México}
\affiliation[ab]{Tsung-Dao Lee Institute \& School of Physics and Astronomy, Shanghai Jiao Tong University, 800 Dongchuan Rd, Shanghai 200240, China}
\affiliation[ac]{Department of Physics, Missouri University of Science and Technology, Rolla, MO, USA}
\affiliation[ad]{Department of Physics, Pennsylvania State University, University Park, PA, USA}

\keywords{axions, gamma ray experiments, dark matter experiments}

\note[*]{Corresponding author.} 
\emailAdd{yoba\_m\_t\_a@ciencias.unam.mx}
\emailAdd{shkdna@sjtu.edu.cn}
\emailAdd{ruben@fisica.unam.mx}

\abstract{
Axion-like particles (ALPs) are hypothetical particles and compelling candidates for cold dark matter. Their existence could be probed through their conversions into photons in the presence of magnetic fields. In this work, we explore the effect of these photon-ALP conversions by searching for an attenuation in the observed gamma ray spectra of galactic sources that emit at energies of hundreds of TeV. We analyze data from the High-Altitude Water Cherenkov (HAWC) Observatory for the source 3HWC J1908+063. No evidence of photon--ALP conversions was found, and we set constraints on the ALP parameter space. Specifically, we derive exclusion limits for ALPs with masses in the range \(10^{-8}~\text{eV} \leq m_a \leq 10^{-6}~\text{eV}\) and photon--ALP couplings in the range \(10^{-12}~\text{GeV}^{-1} \leq g_{a\gamma} \leq 10^{-10}~\text{GeV}^{-1}\), based on HAWC observations.

}

\begin{document}
\maketitle
\flushbottom

\section{Introduction}
\label{sec:intro}

Axions are hypothetical particles proposed as a solution to the strong CP problem in quantum chromodynamics (QCD). This problem concerns the observed conservation of charge–parity (CP) symmetry, even though QCD permits CP violation through a nonzero $\theta$-term. One of the most prominent theoretical resolutions is the Peccei–Quinn mechanism, which introduces a new global $U(1)$ symmetry whose spontaneous breaking predicts the existence of a light pseudoscalar particle known as the axion \cite{PhysRevD.16.1791}. The Peccei–Quinn idea can be extended to broader theoretical frameworks, including string theory, where it naturally predicts a family of particles known as axion-like particles (ALPs) \cite{PhysRevD.37.1237}. Both axions and ALPs are compelling candidates for dark matter due to their potential to exist in a non-relativistic (cold) state, moving at low velocities as predicted by the $\Lambda$CDM cosmological model \cite{Hwang_2009,bauer2018introduction}.

Unlike heavy dark matter candidates such as weakly interacting massive particles (WIMPs), ALPs are characterized by their very low mass, ranging from a few GeV down to $10^{-20}$ eV, which makes their direct detection particularly challenging. However, ALPs possess a unique property: their coupling to electromagnetism. This coupling enables photon-ALP conversions in the presence of magnetic fields, with the conversion probability depending on factors such as energy, magnetic field strength, and propagation distance. Due to this, astrophysical sources represent a significant opportunity for the study of ALPs, as they provide the conditions under which these conversions could occur. In particular, sources that emit at UHE have been of interest because an increase in the photon energy also increases the probability of photon-ALP conversion,  which could manifest as distortions in the observed gamma-ray spectrum \cite{PhysRevLett.40.223, PhysRevLett.40.279,Mirizzi_2017,ABBOTT1983133,MARSH20161,PhysRevLett.51.1415,PhysRevD.84.105030,Galanti:2018upl,DeAngelis:2007dqd,PhysRevLett.99.231102}.

Previous investigations, such as those using Fermi-LAT \cite{PhysRevLett.116.161101},   H.E.S.S \cite{Guo_2021} and High-Altitude Water Cherenkov Observatory (HAWC) public data \cite{Jacobsen_2023}, have focused on distant emission sources, particularly active galaxies like PKS 2155-304 and PG 1553+113, to ensure the presence of these photon-ALP oscillations. However, these studies suffer from systematic uncertainties arising from modeling and limited knowledge of several parameters, such as the extragalactic background light (EBL), magnetic fields, and the intrinsic spectrum \cite{DEANGELIS2008847}.

Advances in instrumentation capable of detecting ultra-high energies \cite{2021Natur.594...33C} have made it possible to study closer galactic sources emitting at tens or hundreds of TeV. Such sources, including pulsar wind nebulae, supernova remnants, and TeV halos, have been reported by HAWC Observatory \cite{Abeysekara_2020}, presenting new opportunities to explore photon-ALP conversions in galactic environments.

Studying galactic sources offers distinct advantages over extragalactic ones, including the avoidance of EBL effects and more accurate modeling of the galactic magnetic field, reducing systematic uncertainties significantly. These factors enable a more reliable interpretation of the high-energy spectra from galactic sources, as uncertainties related to photon propagation through poorly known intergalactic media are minimized \cite{Liang_2019,Zhu_2025,PhysRevD.106.083020,Li_2024}.

The effect of EBL absorption is negligible for galactic sources at energies below PeV due to the relatively short distance between the source and Earth. Additionally, when modeling the magnetic field for a galactic source, only the Milky Way's magnetic field enters the calculations of ALP conversions. However, there are still some important considerations when studying galactic sources. The probability of photon-ALP conversion is dependent on the distance traveled by the photon within the magnetic field, which is considerably reduced in the galactic case compared to the extragalactic scenario. To compensate for this, other parameters in the conversion probability must be increased. This can be achieved by employing this using ultra-high-energy (UHE) photons.

This paper is organized as follows, in Section 2 we present details about ALPs and their mixing with photons, including the conversion probability that is the main observable when considering the effect of oscillations in the photon spectrum of galactic sources. In Section 3, we present the source used in this analysis. Section 4 is dedicated to explaining the analysis methodology used in our work, focusing on the issue of non-nested hypotheses and the approach we use to derive non-biased upper limits. Finally, in Sections 5 and 6 we discuss our results and present our conclusions. In this work, we refer to UHE photons as those detected above tens of TeV and up to a few hundred TeV. Observatories like HAWC, which can detect photons above tens of TeV, are well suited for searching for ALP-induced spectral anomalies, as conversion probabilities become non-negligible in this energy range.

\section{Axion-Like Particles}

ALPs can oscillate into photons in the presence of magnetic fields, an effect expected to be particularly relevant in high-energy astrophysics. The dynamics of ALPs are governed by the Lagrangian \cite{PhysRevD.37.1237}:

\begin{equation}
    \mathcal{L}_{\text{ALP}} = \frac{1}{2}\left(\partial_{\mu} a \, \partial^{\mu} a - m_a^2 \, a^2\right) + \frac{1}{4f_a} a \, F_{\mu \nu} \tilde{F}^{\mu \nu},
\end{equation}

\noindent
where \( a \) represents the ALP field, \( m_a \) is the ALP mass, \( F_{\mu \nu} \) denotes the Faraday tensor, \( \tilde{F}^{\mu \nu} \) is its dual, and \( f_a \) is the ALP decay constant.

The interaction between ALPs and the electromagnetic field is described by:

\begin{equation}
 \mathcal{L}_{a\gamma} = \frac{1}{4f_a} a \, F_{\mu \nu} \tilde{F}^{\mu\nu} = a g_{a\gamma} \, \vec{E} \cdot \vec{B},
\end{equation}

\noindent
where \( g_{a\gamma} = \frac{1}{f_a} \) is the photon-ALP coupling constant, and \( \vec{E} \) and \( \vec{B} \) are the electric and magnetic fields, respectively.

In the presence of an external magnetic field \( \vec{B} \), ALPs and photons can mix, leading to oscillations between these states. This interaction is described by a mixing matrix \(\mathcal{M}\), which encapsulates the equations of motion for the coupled system. Assuming a constant transverse magnetic field aligned with the \( z \)-axis, the mixing matrix for the photon field component \( A_{\parallel} \) (parallel to \( \vec{B} \)) and the ALP field \( a \) can be expressed in Fourier space as:

\begin{equation}
\mathcal{M} =
\begin{pmatrix}
\Delta_{\parallel} & \Delta_{a\gamma} \\
\Delta_{a\gamma} & \Delta_a
\end{pmatrix},
\end{equation}

\noindent
where \( \Delta_{\parallel} \) represents the effective mass term for photons in the medium, \( \Delta_{a\gamma} = g_{a\gamma} B \) is the mixing term that depends on the coupling constant \( g_{a\gamma} \) and the magnetic field strength \( B \), and \( \Delta_a = \frac{m_a^2}{2E_\gamma} \) is the term related to the ALP, with \( m_a \) being the ALP mass and \( E _\gamma\) the photon energy.

To account for the evolution of the quantum state along the direction of propagation, we use the density matrix formalism. The density matrix \(\rho(z)\) describes the quantum state of the system and evolves according to the von Neumann-like equation:

\begin{equation}
i\frac{\mathrm{d}\rho}{\mathrm{d}z} = [\rho,\mathcal{M}].
\end{equation}

The solution to this equation can be expressed using the transfer matrix \(\mathcal{T}(z, 0; E_\gamma)\):

\begin{equation}
\rho(z) = \mathcal{T}(z, 0; E_\gamma) \rho(0) \mathcal{T}^\dagger(z, 0; E_\gamma),
\end{equation}

\noindent
where the transfer matrix \(\mathcal{T}(z, 0; E_\gamma)\) propagates the initial state \(\rho(0)\) over the distance \(z\). This matrix is constructed by considering small steps \(dz\) in the \(z\)-direction, assuming that the magnetic field is constant over each step:

\begin{equation}
\mathcal{T}(z_{N}, z_{1}; E\gamma) = \prod_{i=1}^{N} \mathcal{T}(z_{i+1}, z_{i}; E_\gamma).
\end{equation}

The photon survival probability, \( P_{\gamma\gamma} \), is obtained by tracing the density matrix projected onto the photon states:

\begin{equation}
P_{\gamma\gamma} = \mathrm{Tr}\left( (\rho_{11} + \rho_{22}) \mathcal{T}(z_{N},z_{1}; E_\gamma) \rho(0) \mathcal{T}^\dagger(z_{N},z_{1}; E_\gamma)\right),
\end{equation}

\noindent
here, \( \rho_{11} = \mathrm{diag}(1,0,0) \) and \( \rho_{22} = \mathrm{diag}(0,1,0) \), corresponding to the photon components. This formulation accounts for both the oscillations induced by the magnetic field and variations along the line of sight, allowing for a more precise modeling of photon-ALP conversions and the subsequent evolution of the system.

The photon survival probability can be expressed as:
\begin{equation}\label{prob1}
P_{\gamma\gamma} = \left( 1-P_{\gamma \to a} \right),
\end{equation}
where $P_{\gamma \to a}$ is the photon-ALP conversion probability.

In the case of a homogeneous magnetic field $B$ the probability is given by: 

\begin{equation}\label{probabi}
P_{\gamma \to a}(E_{\gamma}) = \left(1 + \frac{E_c^2}{E_{\gamma}^2}\right)^{-1}\sin^2\left(\frac{g_{a\gamma} B_T L}{2} \sqrt{1 + \frac{E_c^2}{E_{\gamma}^2}}\right),
\end{equation}

\noindent
where \( B_T \) is the transverse magnetic field relative to the photon's direction of motion, \( L \) is the distance traveled within the magnetic field, \( g_{a\gamma} \) is the coupling constant, and \( E_c \) is the critical energy, a reference energy above
which the interaction (mixing) of the photon beam and ALPs becomes significant and can be expressed as:  

\begin{equation}\label{criti}
E_c = \frac{|m_a^2 - \omega_{\text{pl}}^2|}{2g_{a\gamma} B_T},
\end{equation}

\noindent
with \( \omega_{\text{pl}}^2 = \frac{4\pi \alpha n_e}{m_e} \),  the plasma frequency of the medium, \( n_e \) the electron density, \( m_e \) the electron mass, and \( \alpha \) the fine structure constant.

The observed photon flux on Earth from a source, taking into account photon-ALP conversions, is:

\begin{equation}\label{att}
\frac{d\phi}{dE_{\gamma}} = \left(1 - P_{\gamma \to a}\right) f_{\text{att}} \left. \frac{d\phi}{dE_{\gamma}} \right|_{\text{source}}.
\end{equation}

The intrinsic spectrum is denoted as \( \left. \frac{d\phi}{dE_{\gamma}} \right|_{\text{source}} \), and refers to the spectrum emitted at the source. \( P_{\gamma \to a} \) is the photon-ALP conversion probability, and \( f_{\text{att}} \) is a factor that accounts for the attenuation of the flux due to interactions with the EBL and from other effects like local photon fields. For galactic sources, the attenuation factor \( f_{\text{att}} \) is $\approx 1$ due to the proximity of the source to the observer and the energy of the photons. Therefore, the conversion probability \( P_{\gamma \to a} \) is the most relevant factor in the potential distortion of the intrinsic flux.

\section{Source selection and data}

The HAWC Observatory, located at an altitude of 4100m on the slopes of the Sierra Negra volcano in Mexico, is designed to observe gamma rays in the energy range from $\sim$1 TeV up to several hundred TeV. HAWC uses the water cherenkov detection technique, consisting of 300 water cherenkov detectors (WCDs), each instrumented with four photomultiplier tubes that measure the cherenkov light produced by secondary particles in extensive air showers. HAWC operates continuously, covering about two-thirds of the sky every day, and has been taking data since 2015 \cite{ABEYSEKARA2023168253}. The data used in this work correspond to Pass 5 reconstruction which comprises several years of operation and is binned in different energy estimators to cover a wide dynamic range \cite{Albert_2024}. In addition to Galactic and extragalactic source studies, HAWC has performed a variety of dark matter searches. Relevant details can be found in \cite{Albert_2023DM,Abeysekara_2019DM,PhysRevD.109.043034DM,Alfaro_2023} and references therein.

The HAWC Collaboration has reported observations of UHE sources in the 3HWC catalog \cite{Abeysekara_2020}, where multiple galactic sources show emission at energies above 56 TeV. We focus on these sources because the probability of photon-ALP conversion increases with energy, making UHE sources particularly valuable for such studies.

In this work, we focus on the source 3HWC J1908+063, which is one of the brightest galactic sources observed by HAWC at energies exceeding 100 TeV \cite{Albert_2020}. The spectrum of 3HWC J1908+063 has been studied by HAWC and other TeV gamma-ray observatories \cite{Albert_2022,2014ApJ...787..166A,2018A&A...612A...1H}; and it provides an excellent opportunity to search for potential ALP-induced effects.

\begin{figure}[ht]
  \centering
  \begin{subfigure}[b]{0.49\textwidth}
    \includegraphics[width=\textwidth]{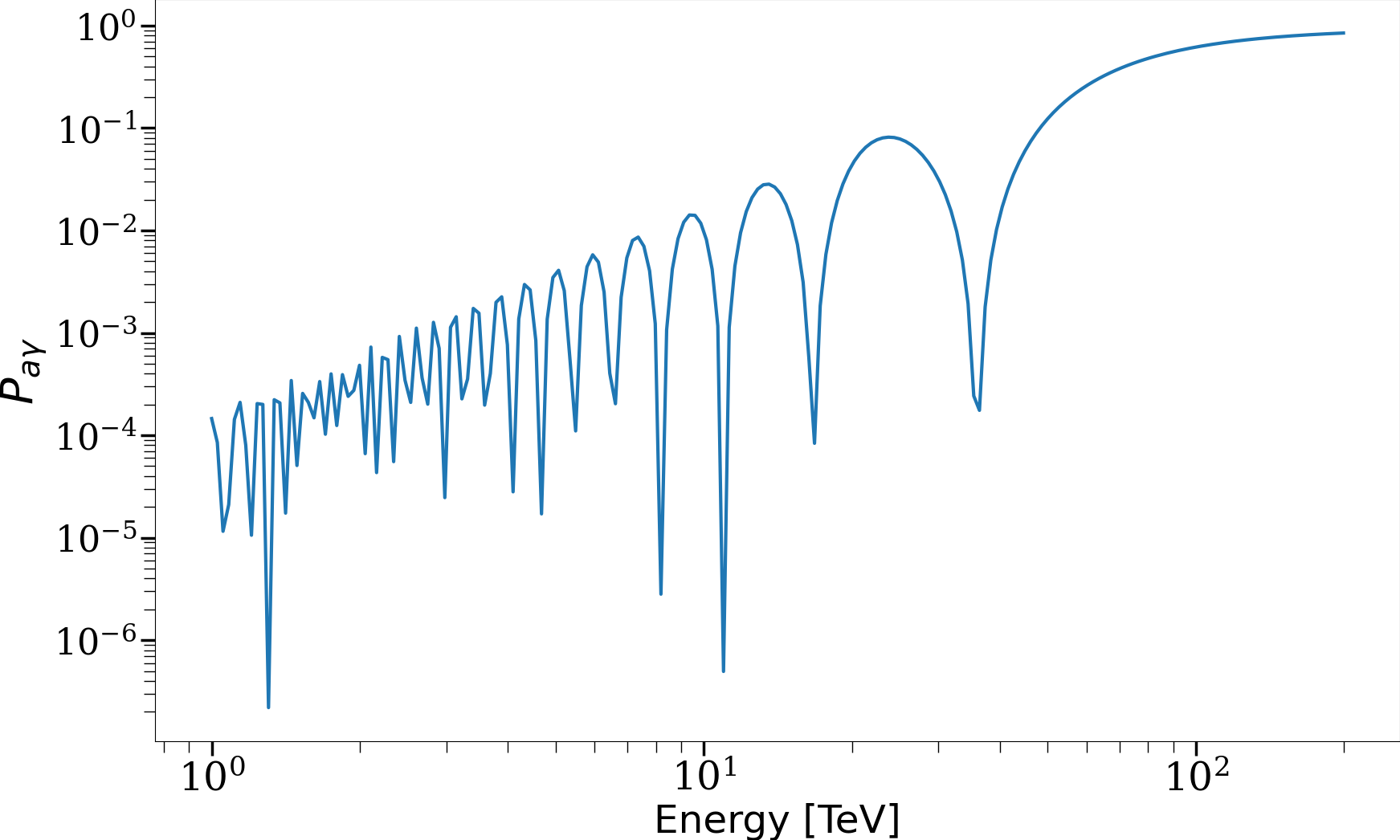}
    
    \label{fig:sub1}
  \end{subfigure}
  \hfill
  \begin{subfigure}[b]{0.49\textwidth}
    \includegraphics[width=\textwidth]{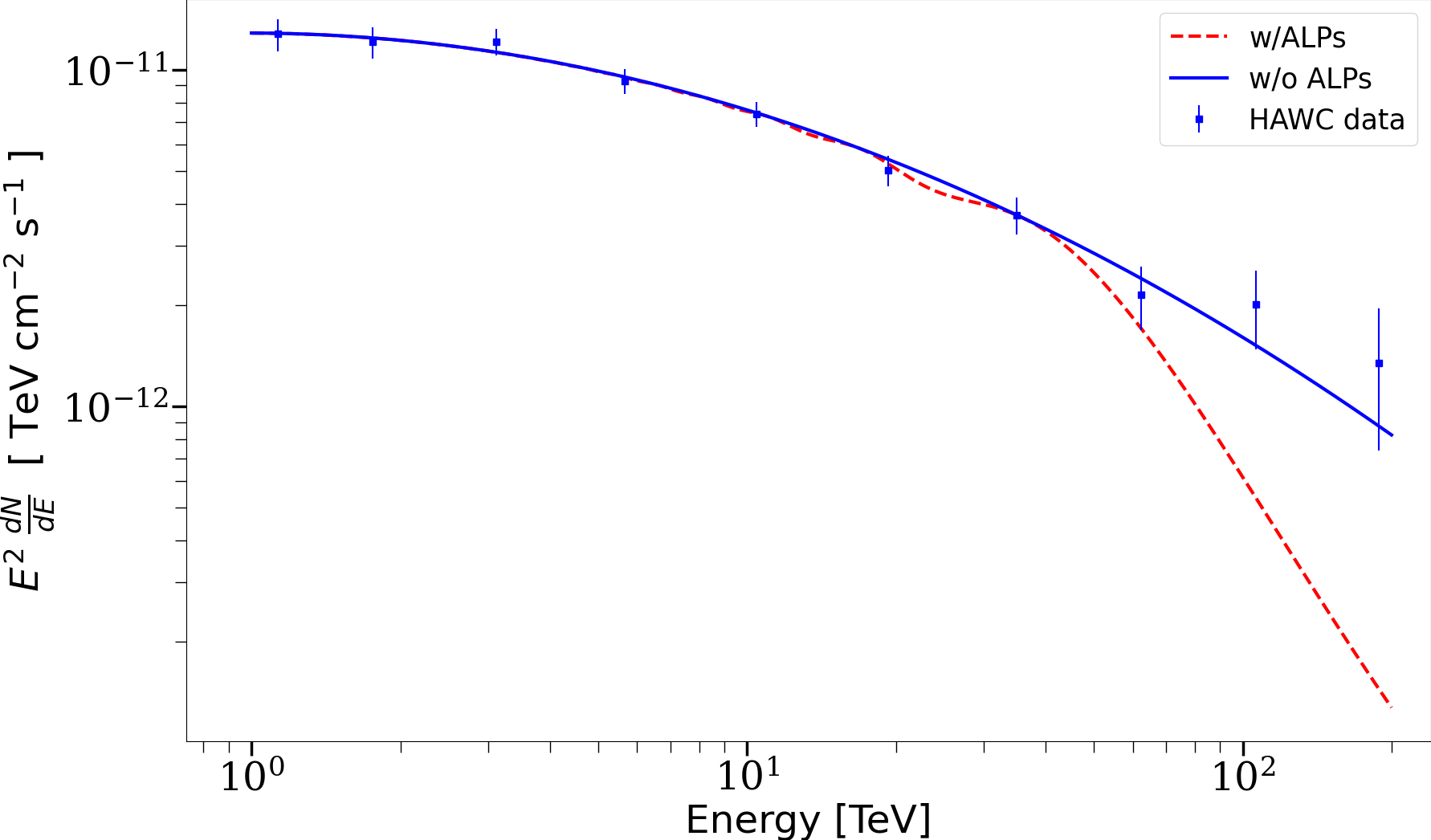}
    
    \label{fig:sub2}
  \end{subfigure}
  \caption{
Left panel: Photon-ALP conversion probability as a function of energy for an ALP candidate with parameters $m_a = 9 \times 10^{-7}~\mathrm{eV}$ and $g_{a\gamma} = 8 \times 10^{-11}~\mathrm{GeV}^{-1}$. 
Right panel: Expected effect of this candidate on the UHE spectrum of the source and  assuming a distance to the earth of 3.2 kpc (red dotted line) compared with the log-parabola fit reported by HAWC without ALPs hypothesis (blue line).
}

  \label{fig:comparacion}
\end{figure}

In order to illustrate and support the claim that UHE gamma rays are ideal for probing ALP-induced effects, it is useful to look at the predicted photon-ALP conversion probabilities for typical galactic parameters (magnetic field strength and distances) in the TeV range. Such a probability curve typically rises with energy, providing a strong motivation to analyze UHE gamma-ray data. Figure \ref{fig:comparacion} shows the variation of the conversion probability for a specific ALP candidate and its potential impact on the source’s gamma-ray spectrum detected by HAWC, particularly at high energies.

The astrophysical parameters associated with the source, such as the distance to Earth and the magnetic field strength, are crucial when calculating the photon survival probability. The source 3HWC~J1909+063 was initially reported by the Milagro observatory and later by H.E.S.S. It has been associated with PSR~J1907+062 and SNR~G40.5$-$0.5 \cite{2007ApJ...664L..91A,2009A&A...499..723A}. Multi-TeV emission can be explained by leptonic processes in the pulsar wind nebula (PWN), with an estimated distance of 3.2~kpc based on dispersion measure (DM) data from Arecibo Observatory. However, multi-TeV emission above 10~TeV can also be explained using lepto-hadronic models connected to the SNR, which has been reported to lie between 3--3.5~kpc and 8--9.5~kpc \cite{DeSarkar_2022}.  This is also one of the motivations for selecting this source: beyond its UHE emission, its distance contributes to enhancing the photon-ALP conversion probability.

Regarding the magnetic field, there are various studies of the Galactic magnetic field. For our analysis, we rely on the model from \cite{Jansson_2012}, which considers distinct components for the disk, halo, and poloidal regions of the Milky Way (MW). However, since the source is located in the Galactic plane, we can assume that the disc is the dominant component and neglect other contributions. As a first approximation, we assume a homogeneous magnetic field of ~3$\mu$G, which is consistent with values reported for the solar neighborhood.

To verify that this approximation is valid, we use the \texttt{GammaALPs} \cite{Meyer:2021pbp} code to calculate the photon survival probability in two different scenarios. In the first scenario, we use the full model from \cite{Jansson_2012}, while in the second, we assume a simpler, homogeneous magnetic field along the line of sight between 3HWC~J1908+063 and Earth. We find that both scenarios yield consistent results, at least for the range of mass and coupling constant of interest. This result reveals an additional advantage of using Galactic sources: it is not always necessary to employ the complete magnetic field model, thereby reducing the computational effort. 

Our results are also consistent with other studies that assume the MW magnetic field can be described by both a regular and a turbulent component, and that at distances of a few kpc, only the regular and homogeneous component is dominant. Taking the previous points into account, in this study we examine several scenarios. Our baseline setup assumes a distance of 3.2~kpc (the distance associated with the PWN) and a magnetic field of 3~$\mu$G. We also explore scenarios with distances of 5~kpc and 8~kpc, which could be consistent with the SNR association. Additionally, to illustrate the magnetic field’s impact, we include a scenario with a 1~$\mu$G field strength, which underscores the importance of the magnetic field in our results.

\section{Method}
\label{sec:method}

To investigate the possible effect induced by photon-ALP conversions on the spectrum of the source 3HWC J1908+063, we compare two spectral models to describe the flux detected by HAWC: a log-parabola model without ALP conversions (null hypothesis), and a log-parabola model including the effects of ALP-photon conversions(alternative hypothesis). The choice of using a log-parabola model for the null hypothesis is motivated by its better fit to the data at energies above 100~TeV, and it is also consistent with the spectrum reported in \cite{DeSarkar_2022}.  That means that in \ref{att}, \begin{equation}
    \left. \frac{d\phi}{dE_{\gamma}} \right|_{\text{source}} = K \left( \frac{E}{E_0} \right) ^ {- \alpha - \beta \log{ \left( \frac{E}{E_0} \right) }},
    \label{logp}
\end{equation}where $E_{0}$ represents the pivot energy and for the null hypothesis $P_{\gamma \to a}=0$. For each model, we obtained the best-fit parameters by fitting the observed data. We then used the likelihood ratio as a test statistic (TS) to decide which model better describes the data and to search for evidence of photon-ALP conversions.

For the alternative hypothesis, which includes ALP-photon conversions, we determined the optimal spectral parameters, $\{K, \alpha, \beta, m_{a}, g_{a\gamma}\}$, by fitting the model described by Eq. \ref{att} and the log-parabola model \ref{logp}.

To define the exclusion region, we utilized the test statistic (TS), calculated as the logarithmic likelihood (LL) ratio:

\begin{equation}
TS = -2 \left( \ln \mathcal{L}(\theta_{0}; m_{a}, g_{a\gamma} = 0) - \ln \mathcal{L}(\hat{\theta}_{m_{a}, g_{a\gamma}}) \right),
\label{llr}
\end{equation}

\noindent where $\ln \mathcal{L}(\theta_{0}; m_{a}, g_{a\gamma} = 0)$ represents the LL under the null hypothesis (log-parabola; no ALPs), and $\ln \mathcal{L}(\hat{\theta}_{m_{a}, g_{a\gamma}})$ is the LL under the alternative hypothesis (ALPs; log-parabola + ALPs) for given values of $m_{a}$ and $g_{a\gamma}$.

The analysis was conducted using the Multi-Mission Maximum Likelihood Framework (3ML)  in conjunction with the HAWC Accelerated Likelihood plugin (HAL) \cite{Albert_2024}. Additionally, the construction of the final exclusion regions in the $(m_a, g_{a\gamma})$ space was performed using the ZEBRA software, which allows a systematic scan over the parameter grid. For each fit, we recorded the LL values for the tested models (log-parabola or log-parabola + ALPs), the LL of the background-only (no source) hypothesis, and the corresponding TS. The LL values for the alternative hypothesis were calculated over a grid of $25 \times 25$ points for $m_a$ and $g_{a\gamma}$ in the region $[10^{-8}, 10^{-5}]~\mathrm{eV} \times [10^{-12}, 10^{-10}]~\mathrm{GeV}^{-1}$ respectively. To complement this, we performed an analysis over a larger region of the sky, including nearby sources, to ensure that the intrinsic spectrum obtained for our source was not affected in any way by surrounding emission.

Due to the presence of non-linear dependencies inherent in ALP-photon conversion probabilities and dependencies on other quantities, such as magnetic field strength and electron density, Wilks' theorem does not hold in this case. Furthermore, the assumption that the null hypothesis (log-parabola) and the alternative hypothesis (including ALP-photon conversions) are nested hypotheses is generally invalid. This is because when the ALP mass is zero, it results in a degenerate solution where the coupling constant $g_{a\gamma}$ can take any value. Previous studies on the impact of ALP-photon conversion on the spectra of galactic and extragalactic sources \cite{PhysRevLett.116.161101, Liang_2019, ABE2024101425, Abdalla_2021} have shown that using classical quantiles (2.71 or 5.99) can lead to under coverage issues in the confidence intervals derived, as the statistical test based on the likelihood does not follow a $\chi^{2}$ distribution with degrees of freedom equal to the difference in parameters between the null and alternative hypotheses.

Following the methodology outlined in \cite{Liang_2019}, we generated a set of pseudo-experiments  by fluctuating the expected model counts for the gamma-ray flux, assuming the null hypothesis (log-parabola). For each PE, we computed likelihoods for the null and the alternative and obtained the TS for the inclusion of the ALP hypothesis against the log-parabola spectrum model. All spectral parameters were allowed to vary freely in each fit. We computed the TS for a total of $300$ pseudo-experiments  and checked whether the resulting distribution could be described by a $\chi^{2}$ distribution. For this distribution, we estimated a new TS threshold to set a 95\% confidence level (C.L.) upper limit for the parameter space $(m_{a}, g_{a\gamma})$. This distribution enables an accurate estimation of the significance of a (non)-detection.

\section{Results}

We present the results obtained from our search for evidence of photon-ALP conversions at TeV energies. First, we present the TS distribution obtained from our pseudo-experiments. Then, as we do not find evidence of ALP-induced effects in the spectrum of 3HWC J1908+063, we estimate the exclusion region in the ALP parameter space.

\subsection{Null TS Distribution}
\label{subsec:nullts_res}

\begin{figure}[ht]
    \centering
    \includegraphics[width=0.8\linewidth]{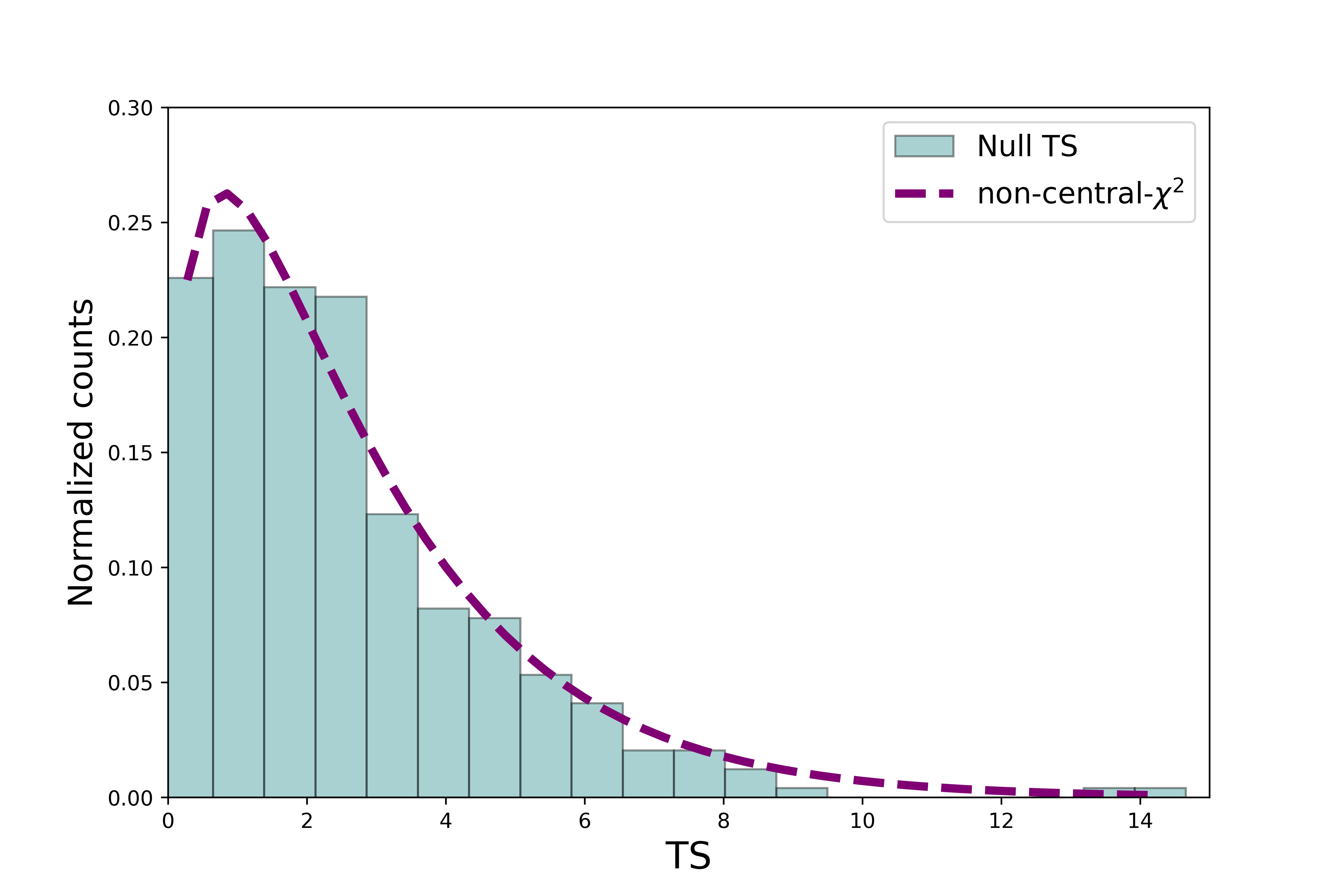}
    \caption{Best fit to the null TS distribution using a $\chi^2$ distribution with effective degrees of freedom.}
    \label{fig:nullts_fit}
\end{figure}

We computed the TS for the ALP hypothesis relative to the null hypothesis. The resulting normalized TS distribution is shown in Figure~\ref{fig:nullts_fit}.

We fitted a $\chi^2$ function to this distribution and found that the distribution is well described by a $\chi^2$ distribution with 2.8 $\pm$ 0.1 effective degrees of freedom. This differs from the expected result of a $\chi^2$ distribution with two degrees of freedom. As previously discussed in Section 4, this deviation arises due to the non-applicability of Wilks' theorem and the non-nested nature of the hypotheses. Using the best-fit parameters of the empirical $\chi^2$ distribution derived from the pseudo-experiments, we determined that the test statistic (TS) value corresponding to the 95\%   confidence level is 7.5. This value was used to establish the exclusion region in the parameter space $(m_{a}, g_{a\gamma})$. Additionally, the TS obtained from the observations of J1908+063 was 4.39, corresponding to an approximately 80\% C.L, indicating that the observed data is consistent with the null hypothesis.. Then, the observed flux of J1908+063 is well described by a log-parabola spectrum, with no statistically significant indication of ALP-induced effects.

\subsection{Exclusion Region}

\begin{figure}[ht!]
    \centering
    \includegraphics[width=0.85\linewidth]{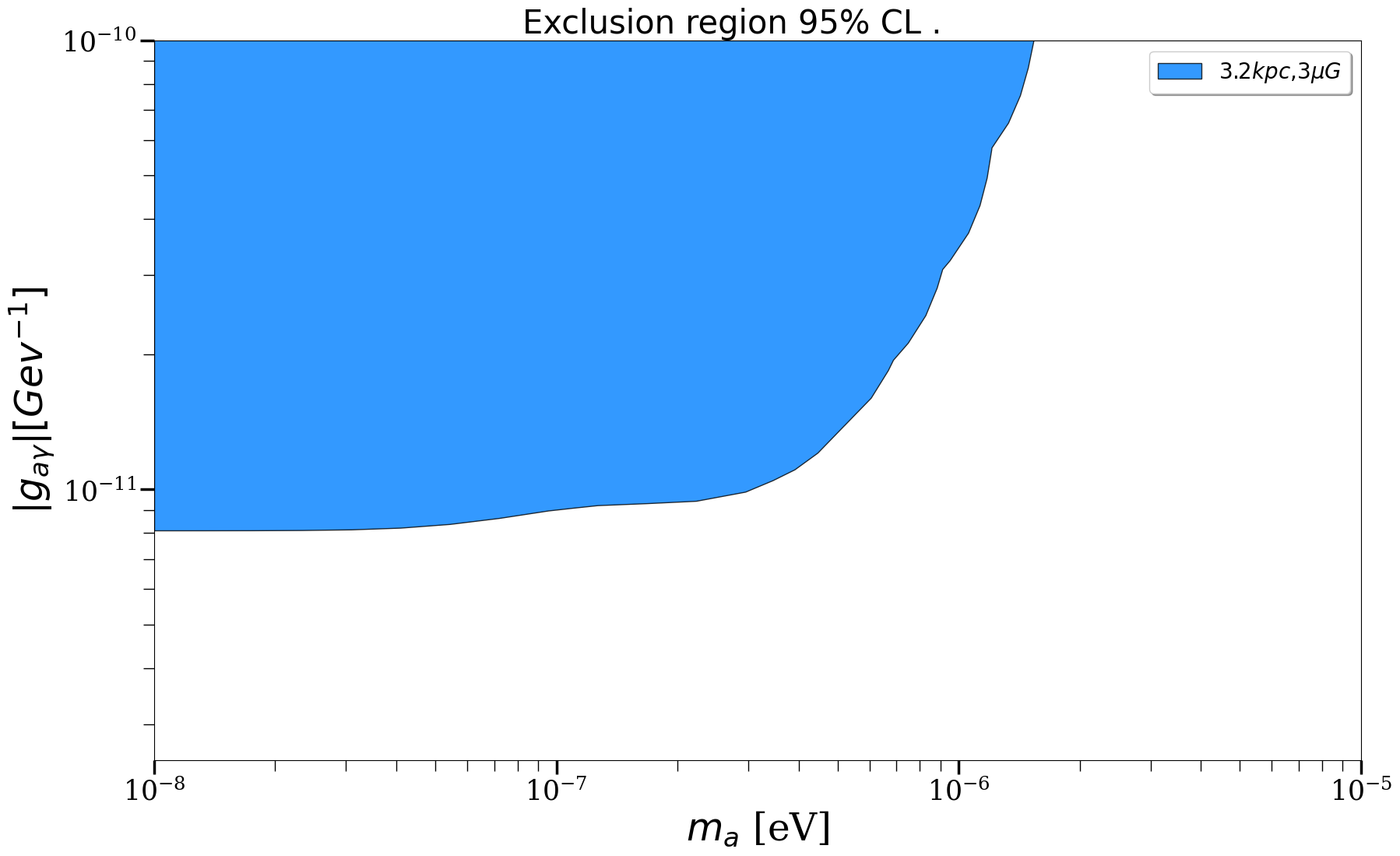}
    \caption{Exclusion region obtained for $m_a$ and $g_{a\gamma}$ based on observations of the galactic source J1908+063, using HAWC data.}
    \label{fig:fhitregion}
\end{figure}

Using the results obtained from the mapping of the likelihood ratio outlined in Section ~\ref{sec:method} and the results from the pseudo-experiments  \ref{subsec:nullts_res}, we determined the exclusion region by applying the $\Delta$TS criteria with a threshold value of 7.5 at the 95\% C.L. The resulting exclusion region is shown in Figure \ref{fig:fhitregion}

\begin{figure}[h!]
    \centering
    \includegraphics
    [width=0.85\linewidth]{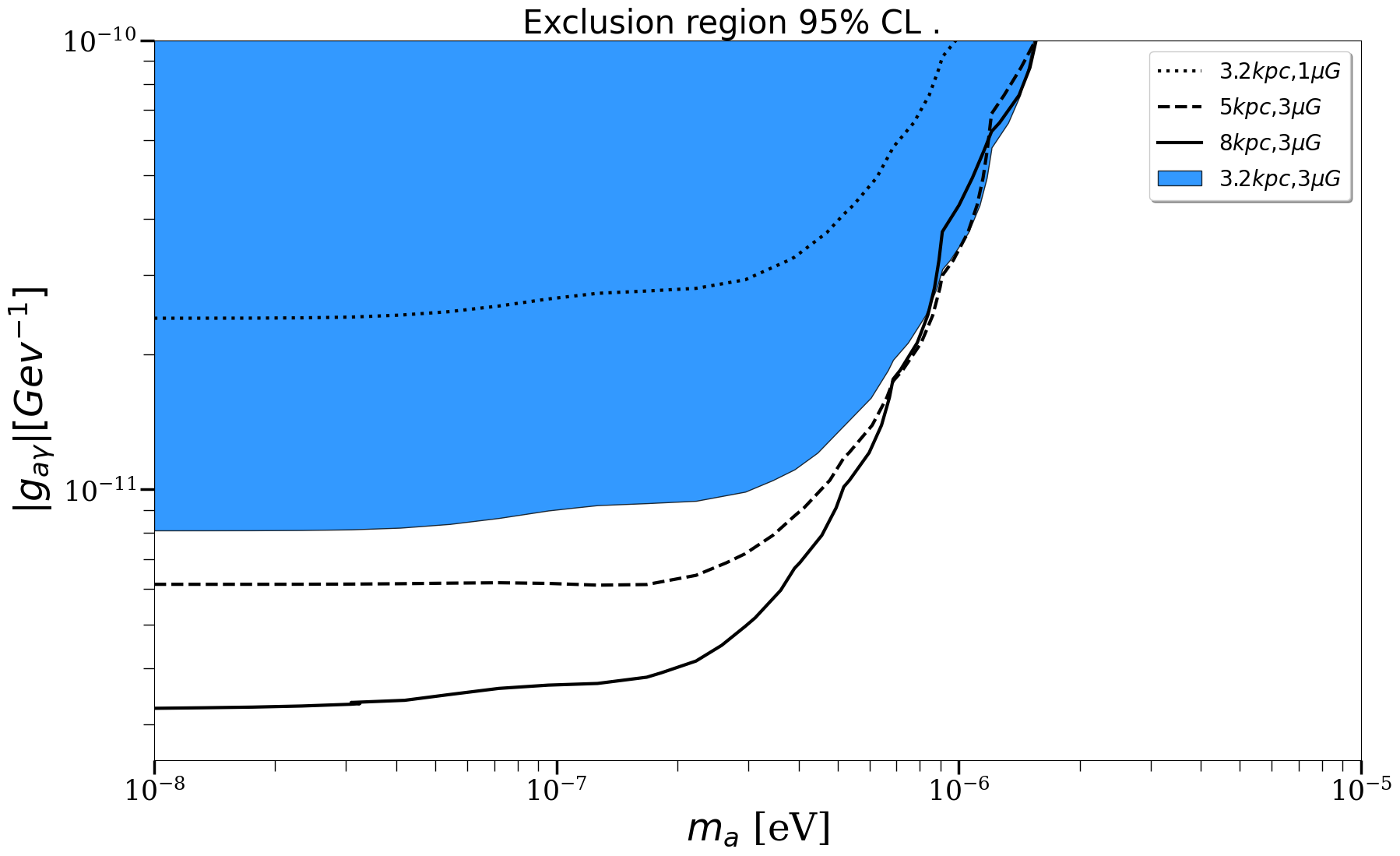}
    \caption{Comparison of exclusion regions for different values of magnetic field strength and distance to the source.}
    \label{fig:fhitregion2}
\end{figure}

To assess the impact of varying parameters on the exclusion region, we examined how changes in distance and magnetic field strength affect the results. The values used for the magnetic field and distance are summarized in Table~\ref{tab:nuisance_params}. The comparison between exclusion regions under different assumptions is shown in Figure~\ref{fig:fhitregion2}, illustrating that for a smaller value of the specified field strength, the exclusion region is reduced. This is consistent with the fact that the conversion probability is proportional to $B_{T}^2$. A similar effect is observed when varying the distance to the source; a larger distance increases the conversion probability due to the longer path length in the magnetic field, expanding the exclusion region.

It can also be observed that the expansion of the exclusion region occurs almost entirely in terms of the coupling constant rather than the mass. To understand this, it is helpful to recall the expression for the critical energy, eq \ref{criti}, which depends inversely on the magnetic field. Consequently, a stronger magnetic field lowers the critical energy, thereby increasing the conversion probability.

In contrast, the dependence of the critical energy on the mass is quadratic. Thus, as the mass increases, the critical energy required to observe conversions grows as the square of the mass. For instance, consider a candidate with $g_{a\gamma} = 8 \times 10^{-12}    \mathrm{ \ GeV^{-1}}$, $m_a = 2 \times 10^{-7}\mathrm{eV}$, and a magnetic field of 1~$\mu\text{G}$; in this scenario, the critical energy is approximately 120~TeV. If the magnetic field alone is raised to 3~$\mu$G for this same candidate, the critical energy drops to about 42~TeV. However, if the mass is also increased to $2 \times 10^{-6} \mathrm{eV}$, the critical energy climbs to nearly 4~PeV which exceeds the current observational energy range. This example underscores the distinct role of the mass, highlighting that exploring candidates with larger masses requires access to much higher energies.

\begin{table}[h!]
    \centering
    \begin{tabular}{|c|c|c|}
    \hline
    Scenario & Magnetic Field ($\mu$G) & Distance (kpc) \\
    \hline
    A & 1 & 3.2 \\
    B & 3 & 3.2 \\
    C & 3 & 5 \\
    D & 3 & 8 \\
    \hline
    \end{tabular}
    \caption{Values of magnetic field strength and distance used to assess the impact of the parameters on the exclusion region. The different parameters were established by taking into account the considerations presented in Section~3.
}
    \label{tab:nuisance_params}
\end{table}

\begin{figure}[h!]
    \centering
    \includegraphics[width=0.85\linewidth]{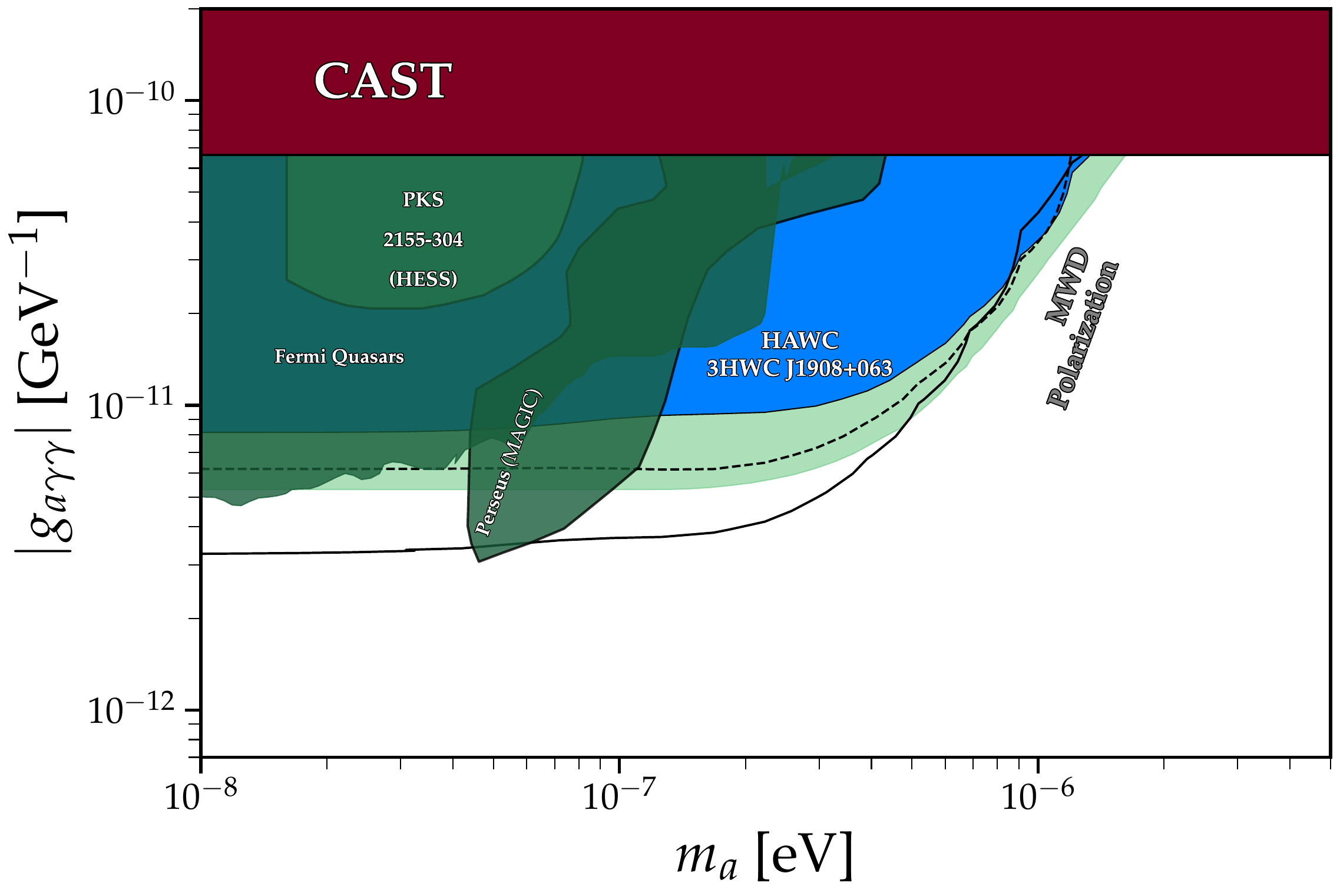}
    \caption{Exclusion regions obtained for different parameter sets are compared with existing results from various experiments, such as CAST \cite{cite-key}, H.E.S.S\cite{HESS:2013udx}, \textit{Fermi}-LAT\cite{PhysRevLett.116.161101},  \cite{ABE2024101425}, and measurements based on magnetic white dwarf polarization \cite{PhysRevD.105.103034}. The HAWC results with baseline values of distance and magnetic field (see text) are shown as the blue shaded band. The effect of varying the distance and B-field values are shown in dashed and solid lines.
}
    \label{fig:comparison}
\end{figure}
\newpage
\section{Conclusions}

This study focused on investigating the potential impact ALPs on the UHE spectrum of the galactic source 3HWC J1908+063. We did not observe statistically significant deviations attributable to ALPs. However, the analysis enables the placement of competitive constraints on the ALP parameter space.

Exclusion regions were determined by varying key astrophysical parameters, such as magnetic field strength and distance to the source, using HAWC data. We observed that these exclusion regions expand as the magnetic field strength or distance increases. This is consistent with the fact that the conversion probability is proportional to $B_{T}^2$ and increases with the path length in the magnetic field.

Furthermore, we performed the spectral fits over an extended sky region that included not only our source 3HWC J1908+063 but also nearby sources, in order to evaluate whether the parameter values changed in any significant way. This analysis confirmed that the presence of nearby sources does not significantly impact the spectral modeling, suggesting that any potential deviations in the spectrum are more likely to be attributable to photon-ALP conversions than to contamination from nearby sources.

The constraints derived from this study contribute to the broader effort to probing the ALP parameter space and demonstrate the utility of UHE observations in setting limits on ALP properties. Future observations with more sensitive instruments and further refinement of the models could provide even tighter constraints or potentially reveal evidence of ALP-induced phenomena. In addition, it would be valuable to compare these limits with those obtained by other observatories (see Figure \ref{fig:comparison}) and methods, thereby situating them within the broader landscape of ALP searches.

\acknowledgments

We acknowledge the support from: the US National Science Foundation (NSF); the US Department of Energy Office of High-Energy Physics; the Laboratory Directed Research and Development (LDRD) program of Los Alamos National Laboratory; Consejo Nacional de Ciencia y Tecnología (CONACyT), México, grants 271051, 232656, 260378, 179588, 254964, 258865, 243290, 132197, A1-S-46288, A1-S-22784, CF-2023-I-645, cátedras 873, 1563, 341, 323, Red HAWC, México; DGAPA-UNAM grants IG101323, IN111716-3, IN111419, IA102019, IN106521, IN110621 , IN110521 , IN102223; VIEP-BUAP; PIFI 2012, 2013, PROFOCIE 2014, 2015; the University of Wisconsin Alumni Research Foundation; the Institute of Geophysics, Planetary Physics, and Signatures at Los Alamos National Laboratory; Polish Science Centre grant, DEC-2017/27/B/ST9/02272; Coordinación de la Investigación Científica de la Universidad Michoacana; Royal Society - Newton Advanced Fellowship 180385; Generalitat Valenciana, grant CIDEGENT/2018/034; The Program Management Unit for Human Resources \& Institutional Development, Research and Innovation, NXPO (grant number

\bibliographystyle{JHEP}
\bibliography{biblio}

@ARTICLE{PhysRevD.37.1237,
       author = {{Raffelt}, Georg and {Stodolsky}, Leo},
        title = "{Mixing of the photon with low-mass particles}",
      journal = {\prd},
         year = 1988,
        month = mar,
       volume = {37},
       number = {5},
        pages = {1237-1249},
          doi = {10.1103/PhysRevD.37.1237},
       adsurl = {https://ui.adsabs.harvard.edu/abs/1988PhRvD..37.1237R}
}

@article{Mirizzi_2017,
	doi = {10.1088/1475-7516/2009/12/004},
  
	url = {https://doi.org/10.1088%2F1475-7516%2F2009%2F12%2F004},
  
	year = 2017,
	month = {jan},
  
	publisher = {{IOP} Publishing},
  
	volume = {2009},
  
	number = {12},
  
	pages = {004--004},
  
	author = {Alessandro Mirizzi and Daniele Montanino},
  
	title = {Stochastic conversions of {TeV} photons into axion-like particles in extragalactic magnetic fields},
  
	journal = {Journal of Cosmology and Astroparticle Physics}
}

@book{bauer2018introduction,
title = {Yet Another Introduction to Dark Matter [E-Book] : The Particle Physics Approach},
series = {Lecture Notes in Physics ;},
author = {Bauer, Martin},
address = {Cham},
publisher = {Springer},
year = {2019},
edition = {1st edition 2019.},
isbn = {9783030162344},
pages = {X, 180 pages 43 illustrations, 22 illustrations in color (online resource)},
note = {English},
url = {https://doi.org/10.1007/978-3-030-16234-4}
}

@article{Hwang_2009,
	author = {Jai-chan Hwang and Hyerim Noh},
	doi = {https://doi.org/10.1016/j.physletb.2009.08.031},
	issn = {0370-2693},
	journal = {Physics Letters B},
	number = {1},
	pages = {1-3},
	title = {Axion as a cold dark matter candidate},
	url = {https://www.sciencedirect.com/science/article/pii/S0370269309009691},
	volume = {680},
	year = {2009}}

@article{PhysRevD.84.105030,
  title = {Relevance of axionlike particles for very-high-energy astrophysics},
  author = {De Angelis, Alessandro and Galanti, Giorgio and Roncadelli, Marco},
  journal = {Phys. Rev. D},
  volume = {84},
  issue = {10},
  pages = {105030},
  numpages = {37},
  year = {2011},
  month = {Nov},
  publisher = {American Physical Society},
  doi = {10.1103/PhysRevD.84.105030},
  url = {https://link.aps.org/doi/10.1103/PhysRevD.84.105030}
}

@ARTICLE{2021Natur.594...33C,
       author = {Cao, Zhen and Aharonian,F.A},
        title = "{Ultrahigh-energy photons up to 1.4 petaelectronvolts from 12 {\ensuremath{\gamma}}-ray Galactic sources}",
      journal = {Nature},
         year = 2021,
        month = jan,
       volume = {594},
       number = {7861},
        pages = {33-36},
          doi = {10.1038/s41586-021-03498-z},
       adsurl = {https://ui.adsabs.harvard.edu/abs/2021Natur.594...33C}
}

@article{Guo_2021,
   title={Implications of axion-like particles from the Fermi-LAT and H.E.S.S. observations of PG 1553+113 and PKS 2155-304 },
   volume={45},
   ISSN={2058-6132},
   url={http://dx.doi.org/10.1088/1674-1137/abcd2e},
   DOI={10.1088/1674-1137/abcd2e},
   number={2},
   journal={Chinese Physics C},
   publisher={IOP Publishing},
   author={Guo, Jun-Guang and Li, Hai-Jun and Bi, Xiao-Jun and Lin, Su-Jie and Yin, Peng-Fei},
   year={2021},
   month={Feb},
   pages={025105}
}

@article{Abeysekara_2020,
   title={Multiple Galactic Sources with Emission Above 56 TeV Detected by HAWC},
   volume={124},
   ISSN={1079-7114},
   url={http://dx.doi.org/10.1103/PhysRevLett.124.021102},
   DOI={10.1103/physrevlett.124.021102},
   number={2},
   journal={Physical Review Letters},
   publisher={American Physical Society (APS)},
   author={Abeysekara, A.U. and Albert, A. and Alfaro, R. and Angeles Camacho, J.R. and Arteaga-Velázquez, J.C. and Arunbabu, K.P. and Avila Rojas, D. and Ayala Solares, H.A. and Baghmanyan, V. and Belmont-Moreno, E. and et al.},
   year={2020},
   month={Jan}
}

@article{Albert_2022,
doi = {10.3847/1538-4357/ac56e5},
url = {https://dx.doi.org/10.3847/1538-4357/ac56e5},
year = {2022},
month = {mar},
publisher = {The American Astronomical Society},
volume = {928},
number = {2},
pages = {116},
author = {A. Albert and R. Alfaro and C. Alvarez and J. D. Álvarez and J. R. Angeles Camacho and J. C. Arteaga-Velázquez and D. Avila Rojas and H. A. Ayala Solares and R. Babu and E. Belmont-Moreno and C. Brisbois and K. S. Caballero-Mora and T. Capistrán and A. Carramiñana and S. Casanova and U. Cotti and J. Cotzomi and S. Coutiño de León and E. De la Fuente and C. de León and R. Diaz Hernandez and B. L. Dingus and M. A. DuVernois and M. Durocher and J. C. Díaz-Vélez and K. Engel and C. Espinoza and K. L. Fan and K. Fang and M. Fernández Alonso and N. Fraija and D. Garcia and J. A. García-González and F. Garfias and G. Giacinti and H. Goksu and M. M. González and J. A. Goodman and J. P. Harding and J. Hinton and B. Hona and D. Huang and F. Hueyotl-Zahuantitla and P. Hüntemeyer and A. Iriarte and A. Jardin-Blicq and V. Joshi and S. Kaufmann and D. Kieda and W. H. Lee and J. Lee and H. León Vargas and J. T. Linnemann and A. L. Longinotti and G. Luis-Raya and K. Malone and V. Marandon and O. Martinez and J. Martínez-Castro and J. A. Matthews and P. Miranda-Romagnoli and J. A. Morales-Soto and E. Moreno and M. Mostafá and A. Nayerhoda and L. Nellen and M. Newbold and M. U. Nisa and R. Noriega-Papaqui and L. Olivera-Nieto and N. Omodei and A. Peisker and Y. Pérez Araujo and E. G. Pérez-Pérez and C. D. Rho and D. Rosa-González and H. Salazar and F. Salesa Greus and A. Sandoval and M. Schneider and H. Schoorlemmer and J. Serna-Franco and A. J. Smith and Y. Son and R. W. Springer and O. Tibolla and K. Tollefson and I. Torres and R. Torres-Escobedo and R. Turner and F. Ureña-Mena and L. Villaseñor and X. Wang and I. J. Watson and E. Willox and A. Zepeda and H. Zhou and HAWC Collaboration and M. Breuhaus and H. Li and H. Zhang},
title = {HAWC Study of the Ultra-high-energy Spectrum of MGRO J1908+06},
journal = {The Astrophysical Journal}
}

@ARTICLE{2009A&A...499..723A,
       author = {{Aharonian}, F. and {Akhperjanian}, A.~G. and {Anton}, G. and {Barres de Almeida}, U. and {Bazer-Bachi}, A.~R. and {Becherini}, Y. and {Behera}, B. and {Benbow}, W. and {Bernl{\"o}hr}, K. and {Boisson}, C. and {Bochow}, A. and {Borrel}, V. and {Braun}, I. and {Brion}, E. and {Brucker}, J. and {Brun}, P. and {B{\"u}hler}, R. and {Bulik}, T. and {B{\"u}sching}, I. and {Boutelier}, T. and {Carrigan}, S. and {Chadwick}, P.~M. and {Charbonnier}, A. and {Chaves}, R.~C.~G. and {Cheesebrough}, A. and {Chounet}, L. -M. and {Clapson}, A.~C. and {Coignet}, G. and {Dalton}, M. and {Daniel}, M.~K. and {Degrange}, B. and {Deil}, C. and {Dickinson}, H.~J. and {Djannati-Ata{\"\i}}, A. and {Domainko}, W. and {O'C. Drury}, L. and {Dubois}, F. and {Dubus}, G. and {Dyks}, J. and {Dyrda}, M. and {Egberts}, K. and {Emmanoulopoulos}, D. and {Espigat}, P. and {Farnier}, C. and {Feinstein}, F. and {Fiasson}, A. and {F{\"o}rster}, A. and {Fontaine}, G. and {F{\"u}{\ss}ling}, M. and {Gabici}, S. and {Gallant}, Y.~A. and {G{\'e}rard}, L. and {Giebels}, B. and {Glicenstein}, J.~F. and {Gl{\"u}ck}, B. and {Goret}, P. and {Hauser}, D. and {Hauser}, M. and {Heinz}, S. and {Heinzelmann}, G. and {Henri}, G. and {Hermann}, G. and {Hinton}, J.~A. and {Hoffmann}, A. and {Hofmann}, W. and {Holleran}, M. and {Hoppe}, S. and {Horns}, D. and {Jacholkowska}, A. and {de Jager}, O.~C. and {Jung}, I. and {Katarzy{\'n}ski}, K. and {Katz}, U. and {Kaufmann}, S. and {Kendziorra}, E. and {Kerschhaggl}, M. and {Khangulyan}, D. and {Kh{\'e}lifi}, B. and {Keogh}, D. and {Komin}, Nu. and {Kosack}, K. and {Lamanna}, G. and {Lenain}, J. -P. and {Lohse}, T. and {Marandon}, V. and {Martin}, J.~M. and {Martineau-Huynh}, O. and {Marcowith}, A. and {Maurin}, D. and {McComb}, T.~J.~L. and {Medina}, M.~C. and {Moderski}, R. and {Moulin}, E. and {Naumann-Godo}, M. and {de Naurois}, M. and {Nedbal}, D. and {Nekrassov}, D. and {Niemiec}, J. and {Nolan}, S.~J. and {Ohm}, S. and {Olive}, J. -F. and {de O{\~n}a Wilhelmi}, E. and {Orford}, K.~J. and {Ostrowski}, M. and {Panter}, M. and {Paz Arribas}, M. and {Pedaletti}, G. and {Pelletier}, G. and {Petrucci}, P. -O. and {Pita}, S. and {P{\"u}hlhofer}, G. and {Punch}, M. and {Quirrenbach}, A. and {Raubenheimer}, B.~C. and {Raue}, M. and {Rayner}, S.~M. and {Renaud}, M. and {Reimer}, O. and {Rieger}, F. and {Ripken}, J. and {Rob}, L. and {Rosier-Lees}, S. and {Rowell}, G. and {Rudak}, B. and {Rulten}, C.~B. and {Ruppel}, J. and {Sahakian}, V. and {Santangelo}, A. and {Schlickeiser}, R. and {Sch{\"o}ck}, F.~M. and {Schr{\"o}der}, R. and {Schwanke}, U. and {Schwarzburg}, S. and {Schwemmer}, S. and {Shalchi}, A. and {Skilton}, J.~L. and {Sol}, H. and {Spangler}, D. and {Stawarz}, {\L}. and {Steenkamp}, R. and {Stegmann}, C. and {Superina}, G. and {Tam}, P.~H. and {Tavernet}, J. -P. and {Terrier}, R. and {Tibolla}, O. and {van Eldik}, C. and {Vasileiadis}, G. and {Venter}, C. and {Venter}, L. and {Vialle}, J.~P. and {Vincent}, P. and {Vivier}, M. and {V{\"o}lk}, H.~J. and {Volpe}, F. and {Wagner}, S.~J. and {Ward}, M. and {Zdziarski}, A.~A. and {Zech}, A.},
        title = "{Detection of very high energy radiation from HESS J1908+063 confirms the Milagro unidentified source MGRO J1908+06}",
      journal = {Astronomy and Astrophysic},
         year = 2009,
        month = jun,
       volume = {499},
       number = {3},
        pages = {723-728},
          doi = {10.1051/0004-6361/200811357},
archivePrefix = {arXiv},
       eprint = {0904.3409},
 primaryClass = {astro-ph.HE},
       adsurl = {https://ui.adsabs.harvard.edu/abs/2009A&A...499..723A}
}

@ARTICLE{2007ApJ...664L..91A,
       author = {{Abdo}, A.~A. and {Allen}, B. and {Berley}, D. and {Casanova}, S. and {Chen}, C. and {Coyne}, D.~G. and {Dingus}, B.~L. and {Ellsworth}, R.~W. and {Fleysher}, L. and {Fleysher}, R. and {Gonzalez}, M.~M. and {Goodman}, J.~A. and {Hays}, E. and {Hoffman}, C.~M. and {Hopper}, B. and {H{\"u}ntemeyer}, P.~H. and {Kolterman}, B.~E. and {Lansdell}, C.~P. and {Linnemann}, J.~T. and {McEnery}, J.~E. and {Mincer}, A.~I. and {Nemethy}, P. and {Noyes}, D. and {Ryan}, J.~M. and {Saz Parkinson}, P.~M. and {Shoup}, A. and {Sinnis}, G. and {Smith}, A.~J. and {Sullivan}, G.~W. and {Vasileiou}, V. and {Walker}, G.~P. and {Williams}, D.~A. and {Xu}, X.~W. and {Yodh}, G.~B.},
        title = "{TeV Gamma-Ray Sources from a Survey of the Galactic Plane with Milagro}",
      journal = {The Astrophysical Journal},
         year = 2007,
        month = aug,
       volume = {664},
       number = {2},
        pages = {L91-L94},
          doi = {10.1086/520717},
archivePrefix = {arXiv},
       eprint = {0705.0707},
 primaryClass = {astro-ph},
       adsurl = {https://ui.adsabs.harvard.edu/abs/2007ApJ...664L..91A}
}

@ARTICLE{2018A&A...612A...1H,
       author = {{H.~E.~S.~S. Collaboration} and {Abdalla}, H. and {Abramowski}, A. and {Aharonian}, F. and {Ait Benkhali}, F. and {Ang{\"u}ner}, E.~O. and {Arakawa}, M. and {Arrieta}, M. and {Aubert}, P. and {Backes}, M. and {Balzer}, A. and {Barnard}, M. and {Becherini}, Y. and {Becker Tjus}, J. and {Berge}, D. and {Bernhard}, S. and {Bernl{\"o}hr}, K. and {Blackwell}, R. and {B{\"o}ttcher}, M. and {Boisson}, C. and {Bolmont}, J. and {Bonnefoy}, S. and {Bordas}, P. and {Bregeon}, J. and {Brun}, F. and {Brun}, P. and {Bryan}, M. and {B{\"u}chele}, M. and {Bulik}, T. and {Capasso}, M. and {Carrigan}, S. and {Caroff}, S. and {Carosi}, A. and {Casanova}, S. and {Cerruti}, M. and {Chakraborty}, N. and {Chaves}, R.~C.~G. and {Chen}, A. and {Chevalier}, J. and {Colafrancesco}, S. and {Condon}, B. and {Conrad}, J. and {Davids}, I.~D. and {Decock}, J. and {Deil}, C. and {Devin}, J. and {deWilt}, P. and {Dirson}, L. and {Djannati-Ata{\"\i}}, A. and {Domainko}, W. and {Donath}, A. and {Drury}, L.~O. 'C. and {Dutson}, K. and {Dyks}, J. and {Edwards}, T. and {Egberts}, K. and {Eger}, P. and {Emery}, G. and {Ernenwein}, J. -P. and {Eschbach}, S. and {Farnier}, C. and {Fegan}, S. and {Fernandes}, M.~V. and {Fiasson}, A. and {Fontaine}, G. and {F{\"o}rster}, A. and {Funk}, S. and {F{\"u}{\ss}ling}, M. and {Gabici}, S. and {Gallant}, Y.~A. and {Garrigoux}, T. and {Gast}, H. and {Gat{\'e}}, F. and {Giavitto}, G. and {Giebels}, B. and {Glawion}, D. and {Glicenstein}, J.~F. and {Gottschall}, D. and {Grondin}, M. -H. and {Hahn}, J. and {Haupt}, M. and {Hawkes}, J. and {Heinzelmann}, G. and {Henri}, G. and {Hermann}, G. and {Hinton}, J.~A. and {Hofmann}, W. and {Hoischen}, C. and {Holch}, T.~L. and {Holler}, M. and {Horns}, D. and {Ivascenko}, A. and {Iwasaki}, H. and {Jacholkowska}, A. and {Jamrozy}, M. and {Jankowsky}, D. and {Jankowsky}, F. and {Jingo}, M. and {Jouvin}, L. and {Jung-Richardt}, I. and {Kastendieck}, M.~A. and {Katarzy{\'n}ski}, K. and {Katsuragawa}, M. and {Katz}, U. and {Kerszberg}, D. and {Khangulyan}, D. and {Kh{\'e}lifi}, B. and {King}, J. and {Klepser}, S. and {Klochkov}, D. and {Klu{\'z}niak}, W. and {Komin}, Nu. and {Kosack}, K. and {Krakau}, S. and {Kraus}, M. and {Kr{\"u}ger}, P.~P. and {Laffon}, H. and {Lamanna}, G. and {Lau}, J. and {Lees}, J. -P. and {Lefaucheur}, J. and {Lemi{\`e}re}, A. and {Lemoine-Goumard}, M. and {Lenain}, J. -P. and {Leser}, E. and {Lohse}, T. and {Lorentz}, M. and {Liu}, R. and {L{\'o}pez-Coto}, R. and {Lypova}, I. and {Marandon}, V. and {Malyshev}, D. and {Marcowith}, A. and {Mariaud}, C. and {Marx}, R. and {Maurin}, G. and {Maxted}, N. and {Mayer}, M. and {Meintjes}, P.~J. and {Meyer}, M. and {Mitchell}, A.~M.~W. and {Moderski}, R. and {Mohamed}, M. and {Mohrmann}, L. and {Mor{\r{a}}}, K. and {Moulin}, E. and {Murach}, T. and {Nakashima}, S. and {de Naurois}, M. and {Ndiyavala}, H. and {Niederwanger}, F. and {Niemiec}, J. and {Oakes}, L. and {O'Brien}, P. and {Odaka}, H. and {Ohm}, S. and {Ostrowski}, M. and {Oya}, I. and {Padovani}, M. and {Panter}, M. and {Parsons}, R.~D. and {Paz Arribas}, M. and {Pekeur}, N.~W. and {Pelletier}, G. and {Perennes}, C. and {Petrucci}, P. -O. and {Peyaud}, B. and {Piel}, Q. and {Pita}, S. and {Poireau}, V. and {Poon}, H. and {Prokhorov}, D. and {Prokoph}, H. and {P{\"u}hlhofer}, G. and {Punch}, M. and {Quirrenbach}, A. and {Raab}, S. and {Rauth}, R. and {Reimer}, A. and {Reimer}, O. and {Renaud}, M. and {de los Reyes}, R. and {Rieger}, F. and {Rinchiuso}, L. and {Romoli}, C. and {Rowell}, G. and {Rudak}, B. and {Rulten}, C.~B. and {Safi-Harb}, S. and {Sahakian}, V. and {Saito}, S. and {Sanchez}, D.~A. and {Santangelo}, A. and {Sasaki}, M. and {Schandri}, M. and {Schlickeiser}, R. and {Sch{\"u}ssler}, F. and {Schulz}, A. and {Schwanke}, U. and {Schwemmer}, S.},
        title = "{The H.E.S.S. Galactic plane survey}",
      journal = {Astronomy and Astrophysic},
         year = 2018,
        month = apr,
       volume = {612},
          eid = {A1},
        pages = {A1},
          doi = {10.1051/0004-6361/201732098},
archivePrefix = {arXiv},
       eprint = {1804.02432},
 primaryClass = {astro-ph.HE},
       adsurl = {https://ui.adsabs.harvard.edu/abs/2018A&A...612A...1H}
}

@ARTICLE{2014ApJ...787..166A,
       author = {{Aliu}, E. and {Archambault}, S. and {Aune}, T. and {Behera}, B. and {Beilicke}, M. and {Benbow}, W. and {Berger}, K. and {Bird}, R. and {Buckley}, J.~H. and {Bugaev}, V. and {Cardenzana}, J.~V. and {Cerruti}, M. and {Chen}, X. and {Ciupik}, L. and {Collins-Hughes}, E. and {Connolly}, M.~P. and {Cui}, W. and {Dumm}, J. and {Dwarkadas}, V.~V. and {Errando}, M. and {Falcone}, A. and {Federici}, S. and {Feng}, Q. and {Finley}, J.~P. and {Fleischhack}, H. and {Fortin}, P. and {Fortson}, L. and {Furniss}, A. and {Galante}, N. and {Gall}, D. and {Gillanders}, G.~H. and {Griffin}, S. and {Griffiths}, S.~T. and {Grube}, J. and {Gyuk}, G. and {Hanna}, D. and {Holder}, J. and {Hughes}, G. and {Humensky}, T.~B. and {Kaaret}, P. and {Kertzman}, M. and {Khassen}, Y. and {Kieda}, D. and {Krennrich}, F. and {Kumar}, S. and {Lang}, M.~J. and {Madhavan}, A.~S. and {Maier}, G. and {McCann}, A.~J. and {Meagher}, K. and {Millis}, J. and {Moriarty}, P. and {Mukherjee}, R. and {Nieto}, D. and {O'Faol{\'a}in de Bhr{\'o}ithe}, A. and {Ong}, R.~A. and {Otte}, A.~N. and {Pandel}, D. and {Park}, N. and {Pohl}, M. and {Popkow}, A. and {Prokoph}, H. and {Quinn}, J. and {Ragan}, K. and {Rajotte}, J. and {Ratliff}, G. and {Reyes}, L.~C. and {Reynolds}, P.~T. and {Richards}, G.~T. and {Roache}, E. and {Rousselle}, J. and {Sembroski}, G.~H. and {Shahinyan}, K. and {Sheidaei}, F. and {Smith}, A.~W. and {Staszak}, D. and {Telezhinsky}, I. and {Tsurusaki}, K. and {Tucci}, J.~V. and {Tyler}, J. and {Varlotta}, A. and {Vassiliev}, V.~V. and {Vincent}, S. and {Wakely}, S.~P. and {Ward}, J.~E. and {Weinstein}, A. and {Welsing}, R. and {Wilhelm}, A.},
        title = "{Investigating the TeV Morphology of MGRO J1908+06 with VERITAS}",
      journal = {The Astrophysical Journal},
         year = 2014,
        month = jun,
       volume = {787},
       number = {2},
          eid = {166},
        pages = {166},
          doi = {10.1088/0004-637X/787/2/166},
archivePrefix = {arXiv},
       eprint = {1404.7185},
 primaryClass = {astro-ph.HE},
       adsurl = {https://ui.adsabs.harvard.edu/abs/2014ApJ...787..166A}
}

@article{Albert_2020,
   title={3HWC: The Third HAWC Catalog of Very-high-energy Gamma-Ray Sources},
   volume={905},
   ISSN={1538-4357},
   url={http://dx.doi.org/10.3847/1538-4357/abc2d8},
   DOI={10.3847/1538-4357/abc2d8},
   number={1},
   journal={The Astrophysical Journal},
   publisher={American Astronomical Society},
   author={Albert, A. and Alfaro, R. and Alvarez, C. and Camacho, J. R. Angeles and Arteaga-Velázquez, J. C. and Arunbabu, K. P. and Rojas, D. Avila and Solares, H. A. Ayala and Baghmanyan, V. and Belmont-Moreno, E. and BenZvi, S. Y. and Brisbois, C. and Caballero-Mora, K. S. and Capistrán, T. and Carramiñana, A. and Casanova, S. and Cotti, U. and de León, S. Coutiño and Fuente, E. De la and Hernandez, R. Diaz and Diaz-Cruz, L. and Dingus, B. L. and DuVernois, M. A. and Durocher, M. and Díaz-Vélez, J. C. and Ellsworth, R. W. and Engel, K. and Espinoza, C. and Fan, K. L. and Fang, K. and Alonso, M. Fernández and Fleischhack, H. and Fraija, N. and Galván-Gámez, A. and Garcia, D. and García-González, J. A. and Garfias, F. and Giacinti, G. and González, M. M. and Goodman, J. A. and Harding, J. P. and Hernandez, S. and Hinton, J. and Hona, B. and Huang, D. and Hueyotl-Zahuantitla, F. and Hüntemeyer, P. and Iriarte, A. and Jardin-Blicq, A. and Joshi, V. and Kieda, D. and Lara, A. and Lee, W. H. and Vargas, H. León and Linnemann, J. T. and Longinotti, A. L. and Luis-Raya, G. and Lundeen, J. and López-Coto, R. and Malone, K. and Marandon, V. and Martinez, O. and Martinez-Castellanos, I. and Martínez-Castro, J. and Matthews, J. A. and Miranda-Romagnoli, P. and Morales-Soto, J. A. and Moreno, E. and Mostafá, M. and Nayerhoda, A. and Nellen, L. and Newbold, M. and Nisa, M. U. and Noriega-Papaqui, R. and Olivera-Nieto, L. and Omodei, N. and Peisker, A. and Araujo, Y. Pérez and Pérez-Pérez, E. G. and Ren, Z. and Rho, C. D. and Rivière, C. and Rosa-González, D. and Ruiz-Velasco, E. and Salazar, H. and Greus, F. Salesa and Sandoval, A. and Schneider, M. and Schoorlemmer, H. and Serna, F. and Sinnis, G. and Smith, A. J. and Springer, R. W. and Surajbali, P. and Tollefson, K. and Torres, I. and Torres-Escobedo, R. and Ukwatta, T. N. and Ureña-Mena, F. and Weisgarber, T. and Werner, F. and Willox, E. and Zepeda, A. and Zhou, H. and León, C. de and Álvarez, J. D.},
   year={2020},
   month=dec, pages={76} }

@article{PhysRevLett.99.231102,
  title = {Detecting Axionlike Particles with Gamma Ray Telescopes},
  author = {Hooper, Dan and Serpico, Pasquale D.},
  journal = {Phys. Rev. Lett.},
  volume = {99},
  issue = {23},
  pages = {231102},
  numpages = {4},
  year = {2007},
  month = {Dec},
  publisher = {American Physical Society},
  doi = {10.1103/PhysRevLett.99.231102},
  url = {https://link.aps.org/doi/10.1103/PhysRevLett.99.231102}
}

@article{DeAngelis:2007dqd,
    author = "De Angelis, Alessandro and Roncadelli, Marco and Mansutti, Oriana",
    title = "{Evidence for a new light spin-zero boson from cosmological gamma-ray propagation?}",
    eprint = "0707.4312",
    archivePrefix = "arXiv",
    primaryClass = "astro-ph",
    doi = "10.1103/PhysRevD.76.121301",
    journal = "Phys. Rev. D",
    volume = "76",
    pages = "121301",
    year = "2007"
}

@article{Meyer:2021pbp,
    author = "Meyer, Manuel and Davies, James and Kuhlmann, Julian",
    title = "{gammaALPs: An open-source python package for computing photon-axion-like-particle oscillations in astrophysical environments}",
    eprint = "2108.02061",
    archivePrefix = "arXiv",
    primaryClass = "astro-ph.HE",
    doi = "10.22323/1.395.0557",
    journal = "PoS",
    volume = "ICRC2021",
    pages = "557",
    year = "2021"
}

@article{Galanti:2018upl,
    author = "Galanti, Giorgio and Tavecchio, Fabrizio and Roncadelli, Marco and Evoli, Carmelo",
    title = "{Blazar VHE spectral alterations induced by photon\textendash{}ALP oscillations}",
    eprint = "1811.03548",
    archivePrefix = "arXiv",
    primaryClass = "astro-ph.HE",
    doi = "10.1093/mnras/stz1144",
    journal = "Mon. Not. Roy. Astron. Soc.",
    volume = "487",
    number = "1",
    pages = "123--132",
    year = "2019"
}

@article{PhysRevD.105.103034,
  title = {Upper limit on the axion-photon coupling from magnetic white dwarf polarization},
  author = {Dessert, Christopher and Dunsky, David and Safdi, Benjamin R.},
  journal = {Phys. Rev. D},
  volume = {105},
  issue = {10},
  pages = {103034},
  numpages = {22},
  year = {2022},
  month = {May},
  publisher = {American Physical Society},
  doi = {10.1103/PhysRevD.105.103034},
  url = {https://link.aps.org/doi/10.1103/PhysRevD.105.103034}
}

@article{PhysRevLett.116.161101,
  title = {Search for Spectral Irregularities due to Photon--Axionlike-Particle Oscillations with the Fermi Large Area Telescope},
  author = {Ajello, M. and Albert, A. and Anderson, B. and Baldini, L. and Barbiellini, G. and Bastieri, D. and Bellazzini, R. and Bissaldi, E. and Blandford, R. D. and Bloom, E. D. and Bonino, R. and Bottacini, E. and Bregeon, J. and Bruel, P. and Buehler, R. and Caliandro, G. A. and Cameron, R. A. and Caragiulo, M. and Caraveo, P. A. and Cecchi, C. and Chekhtman, A. and Ciprini, S. and Cohen-Tanugi, J. and Conrad, J. and Costanza, F. and D'Ammando, F. and de Angelis, A. and de Palma, F. and Desiante, R. and Di Mauro, M. and Di Venere, L. and Dom\'{\i}nguez, A. and Drell, P. S. and Favuzzi, C. and Focke, W. B. and Franckowiak, A. and Fukazawa, Y. and Funk, S. and Fusco, P. and Gargano, F. and Gasparrini, D. and Giglietto, N. and Glanzman, T. and Godfrey, G. and Guiriec, S. and Horan, D. and J\'ohannesson, G. and Katsuragawa, M. and Kensei, S. and Kuss, M. and Larsson, S. and Latronico, L. and Li, J. and Li, L. and Longo, F. and Loparco, F. and Lubrano, P. and Madejski, G. M. and Maldera, S. and Manfreda, A. and Mayer, M. and Mazziotta, M. N. and Meyer, M. and Michelson, P. F. and Mirabal, N. and Mizuno, T. and Monzani, M. E. and Morselli, A. and Moskalenko, I. V. and Murgia, S. and Negro, M. and Nuss, E. and Okada, C. and Orlando, E. and Ormes, J. F. and Paneque, D. and Perkins, J. S. and Pesce-Rollins, M. and Piron, F. and Pivato, G. and Porter, T. A. and Rain\`o, S. and Rando, R. and Razzano, M. and Reimer, A. and S\'anchez-Conde, M. and Sgr\`o, C. and Simone, D. and Siskind, E. J. and Spada, F. and Spandre, G. and Spinelli, P. and Takahashi, H. and Thayer, J. B. and Torres, D. F. and Tosti, G. and Troja, E. and Uchiyama, Y. and Wood, K. S. and Wood, M. and Zaharijas, G. and Zimmer, S.},
  collaboration = {The Fermi-LAT Collaboration},
  journal = {Phys. Rev. Lett.},
  volume = {116},
  issue = {16},
  pages = {161101},
  numpages = {7},
  year = {2016},
  month = {Apr},
  publisher = {American Physical Society},
  doi = {10.1103/PhysRevLett.116.161101},
  url = {https://link.aps.org/doi/10.1103/PhysRevLett.116.161101}
}

@article{Liang_2019,
	author = {Yun-Feng Liang and Cun Zhang and Zi-Qing Xia and Lei Feng and Qiang Yuan and Yi-Zhong Fan},
	doi = {10.1088/1475-7516/2019/06/042},
	journal = {Journal of Cosmology and Astroparticle Physics},
	month = {jun},
	number = {06},
	pages = {042},
	title = {Constraints on axion-like particle properties with TeV gamma-ray observations of Galactic sources},
	url = {https://dx.doi.org/10.1088/1475-7516/2019/06/042},
	volume = {2019},
	year = {2019}
}

@article{ABE2024101425,
	author = {H. Abe and S. Abe and J. Abhir and V.A. Acciari and I. Agudo and T. Aniello and S. Ansoldi and L.A. Antonelli and A. {Arbet Engels} and C. Arcaro and M. Artero and K. Asano and D. Baack and A. Babi{\'c} and A. Baquero and U. {Barres de Almeida} and J.A. Barrio and I. Batkovi{\'c} and J. Baxter and J. {Becerra Gonz{\'a}lez} and W. Bednarek and E. Bernardini and J. Bernete and A. Berti and J. Besenrieder and C. Bigongiari and A. Biland and O. Blanch and G. Bonnoli and {\v Z}. Bo{\v s}njak and I. Burelli and G. Busetto and A. Campoy-Ordaz and A. Carosi and R. Carosi and M. Carretero-Castrillo and A.J. Castro-Tirado and G. Ceribella and Y. Chai and A. Cifuentes and S. Cikota and E. Colombo and J.L. Contreras and J. Cortina and S. Covino and G. D'Amico and V. D'Elia and P. {Da Vela} and F. Dazzi and A. {De Angelis} and B. {De Lotto} and A. {Del Popolo} and J. Delgado and C. {Delgado Mendez} and D. Depaoli and F. {Di Pierro} and L. {Di Venere} and D. {Dominis Prester} and A. Donini and D. Dorner and M. Doro and D. Elsaesser and G. Emery and J. Escudero and L. Fari{\~n}a and A. Fattorini and L. Foffano and L. Font and S. Fukami and Y. Fukazawa and R.J. {Garc{\'\i}a L{\'o}pez} and M. Garczarczyk and S. Gasparyan and M. Gaug and J.G. {Giesbrecht Paiva} and N. Giglietto and F. Giordano and P. Gliwny and N. Godinovi{\'c} and R. Grau and D. Green and J.G. Green and D. Hadasch and A. Hahn and T. Hassan and L. Heckmann and J. Herrera and D. Hrupec and M. H{\"u}tten and R. Imazawa and T. Inada and R. Iotov and K. Ishio and I. {Jim{\'e}nez Mart{\'\i}nez} and J. Jormanainen and D. Kerszberg and G.W. Kluge and Y. Kobayashi and P.M. Kouch and H. Kubo and J. Kushida and M. {L{\'a}inez Lez{\'a}un} and A. Lamastra and F. Leone and E. Lindfors and L. Linhoff and S. Lombardi and F. Longo and R. L{\'o}pez-Coto and M. L{\'o}pez-Moya and A. L{\'o}pez-Oramas and S. Loporchio and A. Lorini and B. {Machado de Oliveira Fraga} and P. Majumdar and M. Makariev and G. Maneva and N. Mang and M. Manganaro and S. Mangano and K. Mannheim and M. Mariotti and M. Mart{\'\i}nez and M. Mart{\'\i}nez-Chicharro and A. Mas-Aguilar and D. Mazin and S. Menchiari and S. Mender and D. Miceli and T. Miener and J.M. Miranda and R. Mirzoyan and M. {Molero Gonz{\'a}lez} and E. Molina and H.A. Mondal and A. Moralejo and D. Morcuende and T. Nakamori and C. Nanci and L. Nava and V. Neustroev and L. Nickel and M. {Nievas Rosillo} and C. Nigro and L. Nikoli{\'c} and K. Nilsson and K. Nishijima and T. {Njoh Ekoume} and K. Noda and S. Nozaki and Y. Ohtani and A. Okumura and J. Otero-Santos and S. Paiano and M. Palatiello and D. Paneque and R. Paoletti and J.M. Paredes and D. Pavlovi{\'c} and M. Persic and M. Pihet and G. Pirola and F. Podobnik and P.G. {Prada Moroni} and E. Prandini and G. Principe and C. Priyadarshi and W. Rhode and M. Rib{\'o} and J. Rico and C. Righi and N. Sahakyan and T. Saito and K. Satalecka and F.G. Saturni and B. Schleicher and K. Schmidt and F. Schmuckermaier and J.L. Schubert and T. Schweizer and A. Sciaccaluga and J. Sitarek and V. Sliusar and D. Sobczynska and A. Spolon and A. Stamerra and J. Stri{\v s}kovi{\'c} and D. Strom and M. Strzys and Y. Suda and S. Suutarinen and H. Tajima and M. Takahashi and R. Takeishi and F. Tavecchio and P. Temnikov and K. Terauchi and T. Terzi{\'c} and M. Teshima and L. Tosti and S. Truzzi and A. Tutone and S. Ubach and J. {van Scherpenberg} and M. {Vazquez Acosta} and S. Ventura and V. Verguilov and I. Viale and C.F. Vigorito and V. Vitale and I. Vovk and R. Walter and M. Will and T. Yamamoto},
	doi = {https://doi.org/10.1016/j.dark.2024.101425},
	issn = {2212-6864},
	journal = {Physics of the Dark Universe},
	pages = {101425},
	title = {Constraints on axion-like particles with the Perseus Galaxy Cluster with MAGIC},
	url = {https://www.sciencedirect.com/science/article/pii/S2212686424000074},
	volume = {44},
	year = {2024}
}

@article{Jansson_2012,
	author = {Jansson, Ronnie and Farrar, Glennys R.},
	journal = {The Astrophysical Journal},
	month = {aug},
	number = {1},
	pages = {14},
	title = {A NEW MODEL OF THE GALACTIC MAGNETIC FIELD},
	volume = {757},
	year = {2012}}

@article{Abdalla_2021,
	author = {H. Abdalla and H. Abe and F. Acero and A. Acharyya and R. Adam and I. Agudo and A. Aguirre-Santaella and R. Alfaro and J. Alfaro and C. Alispach and R. Aloisio and R. Alves Batista and L. Amati and E. Amato and G. Ambrosi and E.O. Ang{\"u}ner and A. Araudo and T. Armstrong and F. Arqueros and L. Arrabito and K. Asano and Y. Ascas{\'\i}bar and M. Ashley and M. Backes and C. Balazs and M. Balbo and B. Balmaverde and A. Baquero Larriva and V. Barbosa Martins and M. Barkov and L. Baroncelli and U. Barres de Almeida and J.A. Barrio and P.-I. Batista and J. Becerra Gonz{\'a}lez and Y. Becherini and G. Beck and J. Becker Tjus and R. Belmont and W. Benbow and E. Bernardini and A. Berti and M. Berton and B. Bertucci and V. Beshley and B. Bi and B. Biasuzzi and A. Biland and E. Bissaldi and J. Biteau and O. Blanch and F. Bocchino and C. Boisson and J. Bolmont and G. Bonanno and L. Bonneau Arbeletche and G. Bonnoli and P. Bordas and E. Bottacini and M. B{\"o}ttcher and V. Bozhilov and J. Bregeon and A. Brill and A.M. Brown and P. Bruno and A. Bruno and A. Bulgarelli and M. Burton and M. Buscemi and A. Caccianiga and R. Cameron and M. Capasso and M. Caprai and A. Caproni and R. Capuzzo-Dolcetta and P. Caraveo and R. Carosi and A. Carosi and S. Casanova and E. Cascone and D. Cauz and K. Cerny and M. Cerruti and P. Chadwick and S. Chaty and A. Chen and M. Chernyakova and G. Chiaro and A. Chiavassa and L. Chytka and V. Conforti and F. Conte and J.L. Contreras and J. Coronado-Blazquez and J. Cortina and A. Costa and H. Costantini and S. Covino and P. Cristofari and O. Cuevas and F. D'Ammando and M.K. Daniel and J. Davies and F. Dazzi and A. De Angelis and M. de Bony de Lavergne and V. De Caprio and R. de C{\'a}ssia dos Anjos and E.M. de Gouveia Dal Pino and B. De Lotto and D. De Martino and M. de Naurois and E. de O{\~n}a Wilhelmi and F. De Palma and V. de Souza and C. Delgado and R. Della Ceca and D. della Volpe and D. Depaoli and T. Di Girolamo and F. Di Pierro and C. D{\'\i}az and C. D{\'\i}az-Bahamondes and S. Diebold and A. Djannati-Ata{\"\i} and A. Dmytriiev and A. Dom{\'\i}nguez and A. Donini and D. Dorner and M. Doro and J. Dournaux and V.V. Dwarkadas and J. Ebr and C. Eckner and S. Einecke and T.R.N. Ekoume and D. Els{\"a}sser and G. Emery and C. Evoli and M. Fairbairn and D. Falceta-Goncalves and S. Fegan and Q. Feng and G. Ferrand and E. Fiandrini and A. Fiasson and V. Fioretti and L. Foffano and M.V. Fonseca and L. Font and G. Fontaine and F.J. Franco and L. Freixas Coromina and S. Fukami and Y. Fukazawa and Y. Fukui and D. Gaggero and G. Galanti and V. Gammaldi and E. Garcia and M. Garczarczyk and D. Gascon and M. Gaug and A. Gent and A. Ghalumyan and G. Ghirlanda and F. Gianotti and M. Giarrusso and G. Giavitto and N. Giglietto and F. Giordano and J. Glicenstein and P. Goldoni and J.M. Gonz{\'a}lez and K. Gourgouliatos and T. Grabarczyk and P. Grandi and J. Granot and D. Grasso and J. Green and J. Grube and O. Gueta and S. Gunji and A. Halim and M. Harvey and T. Hassan Collado and K. Hayashi and M. Heller and S. Hern{\'a}ndez Cadena and O. Hervet and J. Hinton and N. Hiroshima and B. Hnatyk and R. Hnatyk and D. Hoffmann and W. Hofmann and J. Holder and D. Horan and J. H{\"o}randel and P. Horvath and T. Hovatta and M. Hrabovsky and D. Hrupec and G. Hughes and M. H{\"u}tten and M. Iarlori and T. Inada and S. Inoue and A. Insolia and M. Ionica and M. Iori and M. Jacquemont and M. Jamrozy and P. Janecek and I. Jim{\'e}nez Mart{\'\i}nez and W. Jin and I. Jung-Richardt and J. Jurysek and P. Kaaret and V. Karas and S. Karkar and N. Kawanaka and D. Kerszberg and B. Kh{\'e}lifi and R. Kissmann and J. Kn{\"o}dlseder and Y. Kobayashi and K. Kohri and N. Komin and A. Kong and K. Kosack and H. Kubo and N. La Palombara and G. Lamanna and R.G. Lang and J. Lapington and P. Laporte and J. Lefaucheur and M. Lemoine-Goumard and J. Lenain and F. Leone and G. Leto and F. Leuschner and E. Lindfors and S. Lloyd and T. Lohse and S. Lombardi and F. Longo and A. Lopez and M. L{\'o}pez and R. L{\'o}pez-Coto and S. Loporchio and F. Lucarelli and P.L. Luque-Escamilla and E. Lyard and C. Maggio and A. Majczyna and M. Makariev and M. Mallamaci and D. Mandat and G. Maneva and M. Manganaro and G. Manic{\`o} and A. Marcowith and M. Marculewicz and S. Markoff and P. Marquez and J. Mart{\'\i} and O. Martinez and M. Mart{\'\i}nez and G. Mart{\'\i}nez and H. Mart{\'\i}nez-Huerta and G. Maurin and D. Mazin and J.D. Mbarubucyeye and D. Medina Miranda and M. Meyer and S. Micanovic and T. Miener and M. Minev and J.M. Miranda and A. Mitchell and T. Mizuno and B. Mode and R. Moderski and L. Mohrmann and E. Molina and T. Montaruli and A. Moralejo and J. Morales Merino and D. Morcuende-Parrilla and A. Morselli and R. Mukherjee and C. Mundell and T. Murach and H. Muraishi and A. Nagai and T. Nakamori and R. Nemmen and J. Niemiec and D. Nieto and M. Nievas and M. Nikolajuk and K. Nishijima and K. Noda and D. Nosek and S. Nozaki and P. O'Brien and Y. Ohira and M. Ohishi and T. Oka and R.A. Ong and M. Orienti and R. Orito and M. Orlandini and E. Orlando and J.P. Osborne and M. Ostrowski and I. Oya and A. Pagliaro and M. Palatka and D. Paneque and F.R. Pantaleo and J.M. Paredes and N. Parmiggiani and B. Patricelli and L. Pavleti{\'c} and A. Pe'er and M. Pech and M. Pecimotika and M. Peresano and M. Persic and O. Petruk and K. Pfrang and P. Piatteli and E. Pietropaolo and R. Pillera and B. Pilszyk and D. Pimentel and F. Pintore and S. Pita and M. Pohl and V. Poireau and M. Polo and R.R. Prado and J. Prast and G. Principe and N. Produit and H. Prokoph and M. Prouza and H. Przybilski and E. Pueschel and G. P{\"u}hlhofer and M.L. Pumo and M. Punch and F. Queiroz and A. Quirrenbach and R. Rando and S. Razzaque and E. Rebert and S. Recchia and P. Reichherzer and O. Reimer and A. Reimer and Y. Renier and T. Reposeur and W. Rhode and D. Ribeiro and M. Rib{\'o} and T. Richtler and J. Rico and F. Rieger and V. Rizi and J. Rodriguez and G. Rodriguez Fernandez and J.C. Rodriguez Ramirez and J.J. Rodr{\'\i}guez V{\'a}zquez and P. Romano and G. Romeo and M. Roncadelli and J. Rosado and A. Rosales de Leon and G. Rowell and B. Rudak and W. Rujopakarn and F. Russo and I. Sadeh and L. Saha and T. Saito and F. Salesa Greus and D. Sanchez and M. S{\'a}nchez-Conde and P. Sangiorgi and H. Sano and M. Santander and E.M. Santos and A. Sanuy and S. Sarkar and F.G. Saturni and U. Sawangwit and A. Scherer and B. Schleicher and P. Schovanek and F. Schussler and U. Schwanke and E. Sciacca and S. Scuderi and M. Seglar Arroyo and O. Sergijenko and M. Servillat and K. Seweryn and A. Shalchi and P. Sharma and R.C. Shellard and H. Siejkowski and A. Sinha and V. Sliusar and A. Slowikowska and A. Sokolenko and H. Sol and A. Specovius and S. Spencer and D. Spiga and A. Stamerra and S. Stani{\v c} and R. Starling and T. Stolarczyk and U. Straumann and J. Stri{\v s}kovi{\'c} and Y. Suda and P. {\'S}wierk and G. Tagliaferri and H. Takahashi and M. Takahashi and F. Tavecchio and L. Taylor and L.A. Tejedor and P. Temnikov and R. Terrier and T. Terzic and V. Testa and W. Tian and L. Tibaldo and D. Tonev and D.F. Torres and E. Torresi and L. Tosti and N. Tothill and G. Tovmassian and P. Travnicek and S. Truzzi and F. Tuossenel and G. Umana and M. Vacula and V. Vagelli and M. Valentino and B. Vallage and P. Vallania and C. van Eldik and G.S. Varner and V. Vassiliev and M. V{\'a}zquez Acosta and M. Vecchi and J. Veh and S. Vercellone and S. Vergani and V. Verguilov and G.P. Vettolani and A. Viana and C.F. Vigorito and V. Vitale and S. Vorobiov and I. Vovk and T. Vuillaume and S.J. Wagner and R. Walter and J. Watson and M. White and R. White and R. Wiemann and A. Wierzcholska and M. Will and D.A. Williams and R. Wischnewski and A. Wolter and R. Yamazaki and S. Yanagita and L. Yang and T. Yoshikoshi and M. Zacharias and G. Zaharijas and D. Zaric and M. Zavrtanik and D. Zavrtanik and A.A. Zdziarski and A. Zech and H. Zechlin and V.I. Zhdanov and M. {\v Z}ivec},
	doi = {10.1088/1475-7516/2021/02/048},
	journal = {Journal of Cosmology and Astroparticle Physics},
	month = {feb},
	number = {02},
	pages = {048},
	title = {Sensitivity of the Cherenkov Telescope Array for probing cosmology and fundamental physics with gamma-ray propagation},
	url = {https://dx.doi.org/10.1088/1475-7516/2021/02/048},
	volume = {2021},
	year = {2021}
}

@article{PhysRevD.106.083020,
  title = {First constraints on axionlike particles from Galactic sub-PeV gamma rays},
  author = {Eckner, C. and Calore, F.},
  journal = {Phys. Rev. D},
  volume = {106},
  issue = {8},
  pages = {083020},
  numpages = {16},
  year = {2022},
  month = {Oct},
  publisher = {American Physical Society},
  doi = {10.1103/PhysRevD.106.083020},
  url = {https://link.aps.org/doi/10.1103/PhysRevD.106.083020}
}

@article{HESS:2013udx,
    author = "Abramowski, A. and others",
    collaboration = "H.E.S.S.",
    title = "{Constraints on axionlike particles with H.E.S.S. from the irregularity of the PKS 2155-304 energy spectrum}",
    eprint = "1311.3148",
    archivePrefix = "arXiv",
    primaryClass = "astro-ph.HE",
    doi = "10.1103/PhysRevD.88.102003",
    journal = "Phys. Rev. D",
    volume = "88",
    number = "10",
    pages = "102003",
    year = "2013"
}

@article{Jacobsen_2023,
	author = {Jacobsen, Sunniva and Linden, Tim and Freese, Katherine},
	doi = {10.1088/1475-7516/2023/10/009},
	journal = {Journal of Cosmology and Astroparticle Physics},
	month = {oct},
	number = {10},
	pages = {009},
	publisher = {IOP Publishing},
	title = {Constraining axion-like particles with HAWC observations of TeV blazars},
	url = {https://dx.doi.org/10.1088/1475-7516/2023/10/009},
	volume = {2023},
	year = {2023}}
\end{document}